\documentclass[10pt,twocolumn,twoside]{IEEEtran}
\usepackage{array}
\usepackage{cite}
\usepackage{calc}
\usepackage{caption}
\usepackage{graphicx}
\usepackage{psfrag}
\usepackage[tight,footnotesize]{subfigure}
\usepackage{stfloats}
\usepackage{url}
\usepackage{subfig}
\usepackage{longtable}
\usepackage{stfloats}
\usepackage{amsfonts,amssymb,amsbsy,bm,paralist,theorem,cite,ifthen,color}
\usepackage{amsmath}
\usepackage{algorithmic,algorithm}

\newcommand\Xb{\ensuremath{{\bm X}}}

\newcommand\Wb{\ensuremath{{\bm W}}}

\newcommand\lambdab{\ensuremath{{\bm \lambda}}}

\newcommand\E{\ensuremath{{\mathbb{E}}}}

\newcommand\Prob{\ensuremath{{\rm Prob}}}

\newtheorem{Remark}{Remark}

\begin{document}
\bibliographystyle{IEEEtran}
\title{Real-Time Power Balancing via Decentralized Coordinated Home Energy Scheduling}

\author{\vspace{0.5cm} Tsung-Hui Chang, \emph{Member, IEEE}, Mahnoosh Alizadeh, \emph{Student Member, IEEE}, \\and Anna Scaglione, \emph{Fellow, IEEE}
\thanks{Copyright (c) 2013 IEEE. Personal use of this material is permitted. However, permission to use this material for any other purposes must be obtained from the IEEE by sending a request to pubs-permissions@ieee.org.}
\thanks{
Tsung-Hui Chang is the corresponding author. Address:
Department of Electronic and Computer Engineering, National Taiwan University of Science and Technology, Taipei 10607, Taiwan (R.O.C). E-mail: tsunghui.chang@ieee.org.
}
\thanks{
Mahnoosh Alizadeh and Anna Scaglioneand are with Department of Electrical and Computer Engineering, University of California, Davis, CA 95616. E-mail: \{malizadeh,ascaglione\}@ucdavis.edu. }
\vspace{-0.5cm}
\thanks{The work is supported partly by the DOE centers CERTS and TCiPG, and partly by the
National Science Council, Taiwan, under grant NSC 101-2218-E-011-043.}}

\markboth{To appear in IEEE TRANSACTIONS ON Smart Grid, 2013}  \maketitle

\maketitle

\begin{abstract}
It is anticipated that an uncoordinated operation of individual home energy management (HEM) systems in a neighborhood would have a rebound effect on the aggregate demand profile.
To address this issue, this paper proposes a coordinated home energy management (CoHEM) architecture in which distributed HEM units collaborate with each other in order to keep the demand and supply balanced in their neighborhood.
Assuming the energy requests by customers are random in time, we formulate the proposed CoHEM design as a multi-stage stochastic optimization problem. We propose novel models to describe the deferrable appliance load (e.g., Plug-in (Hybrid) Electric Vehicles (PHEV)), and apply approximation and decomposition techniques to handle the considered design problem in a decentralized fashion. The developed decentralized CoHEM algorithm allow the customers to locally compute their scheduling solutions using domestic user information and with message exchange between their neighbors only.
{Extensive simulation results demonstrate that the proposed CoHEM architecture can effectively improve real-time power balancing. Extension to joint power procurement and real-time CoHEM scheduling is also presented.}
\end{abstract}

\IEEEpeerreviewmaketitle
\vspace{-0.3cm}
\section{Introduction}

Demand side management (DSM) techniques have been employed to make the inelastic demand for electricity flexible in order to achieve the goal of integrating intermittent renewable energy resources in the power grid {\cite{dsm1,dsm2}}.
Existing DSM techniques can be divided into {two categories-- Direct load control (DLC) and dynamic pricing. DLC} mechanisms {\cite{dlc1,dlc2}} allow electric utilities to conduct centralized demand management on certain interruptible appliances, e.g., turning off air conditioning systems for short periods of time, to maintain demand and supply balance during peak hours. On the contrary, dynamic pricing schemes {\cite{dp1,dp2}} distribute a control signal to the customers that reflects the congestion of the grid, counting on the assumption that individual customers will modify their demands accordingly.
Since a manual control in response to ever-changing price signals is infeasible, dynamic pricing solutions require intelligent energy management software to manage the demands of residential and commercial customers. Consequently, there is an extensive literature emerging on home energy management (HEM) systems, e.g.,  \cite{han,rad,Du2011,Kim2011}, {dedicated to finding optimal scheduling algorithms for the household appliances,
based on
the price signals, appliance load profiles, job deadlines, etc., with the goal of minimizing the electricity bill.}

As observed in \cite{Kishore2010}, if all the customers in a neighborhood are given the same dynamic price, the HEM systems that are individually operated by each customer will simultaneously schedule the load to the low-price period, and, consequently, a new ``rebound" peak may occur. Another concern is that dynamic prices may render the demand more volatile and less predictable, causing serious stability and reliability issues for the grid if not handled properly \cite{Roozbehani}. In this paper, we aim to blur the boundaries between DLC and dynamic pricing strategies by proposing an architecture through which the HEM units inside the territory of an aggregator/retailer can cooperate with each other to keep wholesale demand of the retailer balanced with the available generation supply (which might be the day-ahead/hour-ahead bid plus locally generated renewable resources). 

{\bf Related works:} Several existing works have studied neighborhood-wise collaborative energy management, though different models and optimization goals are considered.
For example, in \cite{Kishore2010} a heuristic algorithm is proposed for scheduling the load of customers in a neighborhood to meet a maximum power profile specified by the retailer. In \cite{Mohsenian-Rad2010TSG,Mohsenian-Rad2010,Caron2010}, distributed energy management algorithms, based on game-theoretic approaches, were proposed to minimize the cost of the retailer or the peak-to-average ratio of the aggregate load.
The works in \cite{Gatsis2011,Gatsis2011TSG} and \cite{LiLow2011PES} proposed distributed energy management algorithms that maximize the social welfare by minimizing both the costs of retailer and customers. The work in \cite{JiangLow2011CDC} considered simultaneous procurement of power in the day-ahead market and real-time load scheduling to {maximize} the social welfare. However, most of the literature cited above does not account for the customers' probabilistic behavior in the usage of appliances, {which is very important for making decisions that meet the current need of the customers as well as for predicting} what else the customer may want in the future.
Moreover, a common assumption in these papers is that the HEM system can adjust the appliance power consumptions, which may not be valid for some non-interruptible appliances.
{In \cite{ddls2,MahnooshJSAC12}, a neighborhood-level load scheduling architecture is proposed by modulating and scheduling the demand of certain deferrable appliances. This architecture, however, is implemented in a centralized fashion and cannot guarantee the deadline constraints for customers.}

{\bf Contributions:} {In this paper, we propose a coordinated HEM (CoHEM) architecture where the HEM units in a neighborhood collaborate to minimize the cost of their aggregator/retailer in the real-time balancing market. }
Similar to previous works \cite{rad,Du2011,Kim2011,Caron2010}, we assume that the retailer broadcasts dynamic pricing information to the residences, and that the associated HEM systems optimize the scheduling of their local appliances. Unlike \cite{rad,Du2011,Kim2011}, and similar to \cite{Mohsenian-Rad2010TSG,Gatsis2011,Gatsis2011TSG,LiLow2011PES,JiangLow2011CDC}, the CoHEM customers will not be selfish. One of the key differences of our model compared to \cite{Kishore2010,Mohsenian-Rad2010TSG,Mohsenian-Rad2010,Caron2010,Gatsis2011,Gatsis2011TSG,LiLow2011PES,JiangLow2011CDC} is in the choice of the network utility and in the effective pricing and service policy applied to the HEM users. Specifically, in our model, the HEMs, by cooperating, do not pay more than their selfishly optimized cost and do not experience lower quality of service.
The network utility of the CoHEM is chosen to be equal to the cost of deviating in real time from the bulk power purchase of the retailer in the day-ahead and hour-ahead markets, computed using the locational marginal prices \cite{Zheng2006}.

{Another major difference compared to \cite{Mohsenian-Rad2010TSG,Gatsis2011,Gatsis2011TSG,LiLow2011PES,JiangLow2011CDC}, but similar to \cite{Kim2011,Caron2010,Kishore2010},} is that our work focuses on appliances that are flexible in deferring their operating times, e.g., PHEV, washing machine, dish washer and tumble dryer etc. (which are usually non-interruptible in power consumption). Moreover, in order to be foresighted, we take into account the customers' probabilistic behavior by assuming that the customers randomly submit requests to their HEM to use an appliance. Given the statistical information of customers and quality of service constraints (i.e., scheduling deadline constraints), {we formulate the proposed CoHEM design problem as a multi-stage stochastic optimization problem \cite{Shapiro2007,BK:Bersekas07}, aiming at minimizing the expected real-time power unbalancing cost of the retailer. The stochastic formulation can provide
optimal control policies which can be used for real-time appliance scheduling by exploiting both customers' statistical and real-time
request information.

To this end, we first present a Markov decision process (MDP) formulation that can efficiently solve the selfish HEM design problem. Subsequently, we show that the MDP method can be used to develop a \emph{decentralized scheduling algorithm} for efficiently handling the proposed CoHEM design problem.
{It is known that a centralized computation requires full knowledge of all the statistical and real-time information of the customers, and moreover, the required computational complexity increases with the number of residences and the number of controllable appliances. In view of this, a distributed implementation algorithm, that can decompose the original problem into parallel subproblems with smaller problem sizes, is of great interest. Such a distributed algorithm can be deployed in the neighborhood in a fully decentralized fashion where each of the residences computes its scheduling solution locally using domestic information and by communicating with its neighbors only. Since no explicit information about customers' electricity usage is exchanged and submitted to the retailer,
this decentralized method also preserves customers' privacy with respect to their appliance usage.
Extensive simulation results will be presented to demonstrate the effectiveness of the proposed CoHEM architecture and the decentralized scheduling algorithm.}

{\bf Synopsis:} In Section \ref{sec: model and prob}, we first present the load models of deferrable appliances, and the individual (selfish) HEM design problem. Secondly, we present the proposed CoHEM architecture and the associated CoHEM design formulation.
In Section \ref{sec: methods}, an MDP method for solving the selfish HEM design problem is proposed. Then, based on this MDP method and decomposition techniques, we propose a decentralized CoHEM algorithm. Two extensions of the proposed CoHEM design are also discussed in the last subsection. Extensive simulation results are presented in Section \ref{sec: simulations}. Finally, the conclusions and future directions are included in Section \ref{sec: conclusions}.}

\vspace{-0.2cm}
\section{Appliance Load Model and Problem Statement}\label{sec: model and prob}
In the first two subsections, we present the load model of deferrable appliances and the individual HEM problem formulation. The proposed CoHEM architecture is presented in the third subsection.

\vspace{-0.2cm}
\subsection{Multi-Mode Deferrable Appliance Load Model}\label{subsec: load model}

We consider the case where there are $N$ deferrable appliances in each residence. The appliances are assumed to have known power consumption profiles, and once are turned on, their operation cannot be interrupted, e.g., PHEV, dish washer, tumble dryer etc. The HEM system in the house is allowed to defer their schedules within the deadlines specified by the customers. {Specifically, given a request submitted by the customer, the HEM unit has to decide to turn on the appliance immediately or defer the task by waiting in a queue.
The decision process is repeated until the HEM system chooses to activate the appliance or until the maximum delay time is reached.
}

In the paper, we assume that the appliance may have multiple operation modes, each with a different power consumption profile, {and they are decided by the customer. For example, the washing machine may have one mode for colored clothes and one mode for white clothes.}
We assume that appliance $i$ has $M_i$ modes, and each mode specifies a (discrete-time) power load profile
$g_{i,m}(t)$, $t=1,\ldots,G_{i,m}$, where $G_{i,m}>0$ is the maximum job length of $g_{i,m}(t)$. The times at which the customer submits a request for an appliance are random. In particular, 
at each time $t$, appliance $i$ will be requested by the customer with probability $p_i(t)\in [0,1]$, and, once it is requested, it is with probability $\gamma_{i,m}(t)$ that mode $m$ will be chosen ($\sum_{m=1}^{M_i}\gamma_{i,m}(t)=1$), where $t=1,\ldots,T$, with $T>0$ denoting the maximum look-ahead time horizon.
Information about $p_i(t)$ and $\gamma_{i,m}(t)$ can be estimated through the usage history of customers; see, e.g., \cite{Paater2006,Widen2010} for related papers.
Suppose that requests for turning on appliance $i$ arrive at times $t_{i,1}, t_{i,2}, \ldots$ $\in \{1,\ldots,T\}$. Moreover, let $\theta_i(t_{i,k})\in \{1,\ldots,M_i\}$ denote the operation mode chosen at time $t_{i,k}$.
Then, without scheduling, the power load due to appliance $i$ would be
\begin{align}\label{unscheduled load}
   {L}_i(t)&=
   \sum_{k=1}^{\infty} g_{i,\theta_i(t_{i,k})}(t- t_{i,k})~\forall t. 
\end{align}

The requested appliance tasks may be queued and scheduled to operate later. Let $s_{i,1}$, $ s_{i,2}$, $\ldots$ $\in \{1,\ldots,T\}$, be the scheduled times {determined by the HEM system} for turning on appliance $i$, where $s_{i,k} \geq t_{i,k}$ for all $k$. Then, the scheduled power load of appliance $i$ is given by
\begin{align}\label{eq:load injection of appliance i}
   D_i(t)&=\sum_{k=1}^{\infty}g_{i,\theta_i(t_{i,k})}(t- s_{i,k})~\forall t.  
\end{align}  Taking into account the power load of the uncontrollable appliances, denoted by ${U}(t)$, the aggregate power load of a residence can be expressed as
\begin{align}\label{eq:total load}
     L_{{{\rm total}}}(t) = {U}(t) + \sum_{i=1}^N D_{i}(t),~t=1,\ldots,T.
\end{align}

\vspace{-0.4cm}
\subsection{HEM Design Problem}\label{subsec: hem}
Let $\pi(t),$ $t=1,\ldots,T,$ be the dynamic electricity prices given by the retailer.
The HEM targets to schedule the controllable appliances such that the expected total electricity cost of the customer, i.e.,
\begin{align}\label{eq:bill}
   \sum_{t=1}^T \E\{\pi(t)  L_{{{\rm total}}}(t)\}
\end{align}
is minimized, where $\E\{\cdot\}$ denotes the expectation operator. The scheduling task is usually subject to constraints that reflect the customer's degree of comfort. In this work, we assume that the customer will preassign a maximum tolerable delay for each appliance, and that the HEM system 
must not exceed the specified delay. 
In particular, we let $\zeta_{i,m}\geq 0$ denote the maximum delay for mode $m$ of appliance $i$. Then the operating times of appliance $i$ have to satisfy
\begin{align}\label{eq:deadline constraint0}
  t_{i,k} \leq s_{i,k} \leq  t_{i,k}+\zeta_{i,\theta_i(t_{i,k})}~\forall k.
\end{align} 

Mathematically, the HEM design problem can be formulated as the following {\emph{multi-stage stochastic optimization problem}}
\begin{subequations}\label{eq:HEMS}
\begin{align}
  \min_{\substack{s_{i,1},s_{i,2},\ldots \\ \forall~i=1,\ldots,N}}~&\sum_{t=1}^T \E\left\{\pi(t) \left(\sum_{i=1}^N D_{i}(t)+U(t)\right)\right\} \label{eq:HEMS a} \\
  \text{s.t.}~& D_i(t)=\sum_{k=1}^{\infty} g_{i,\theta_i(t_{i,k})}(t- s_{i,k})~\forall i,t, \label{eq:HEMS b}\\
  & t_{i,k} \leq s_{i,k} \leq  t_{i,k}+\zeta_{i,\theta_i(t_{i,k})}~\forall i,k, \label{eq:HEMS c}\\
  & s_{i,k} \leq \max\{T-G_{i,\theta_i(t_{i,k})},t_{i,k}\}~ \forall i,k, \label{eq:HEMS d}
\end{align}
\end{subequations}
where, in \eqref{eq:HEMS a}, the expectation is with respect to the random arrival times $\{t_{i,k}\}$, random operation modes $\{\theta_i(t_{i,k})\}$ and the control variables $\{s_{i,k}\}$, {which are all based on the statistical usage information of the customer. The constraints in
\eqref{eq:HEMS c} and \eqref{eq:HEMS d}, however, specify the real-time scheduling constraints. Specifically, \eqref{eq:HEMS c}
implies that the customer won't wait longer than the specified delay $\zeta_{i,m}$ 
in real time;} while \eqref{eq:HEMS d} implies that all the requested tasks have to be finished
before time $T$ if they arrive before time $T-G_{i,\theta_i(t_{i,k})}$; otherwise, they should be activated for service right after their arrivals\footnote{Note that if \eqref{eq:HEMS d} is not imposed and if $\zeta_{i,\theta_i(t_{i,k})} \geq G_{i,\theta_i(t_{i,k})}$, the scheduler would not turn on the appliance until the end of the horizon $T$, since it contributes no cost to the objective function in \eqref{eq:HEMS a}.}.

{An important aspect for
the multi-stage stochastic formulation \eqref{eq:HEMS} is that, according to the stochastic optimization theory and dynamic programming techniques \cite{BK:Bersekas07,Shapiro2007}, we can find
a so called optimal \emph{control policy}\footnote{By stochastic optimization \cite{BK:Bersekas07,Shapiro2007}, a control policy is a function of the problem states and is like a table that lists all the corresponding actions that the controller should follow for all possible states of the problem. Once a control policy is obtained (e.g., by the dynamic programming techniques \cite{BK:Bersekas07}), the controller can control the appliance in real time by applying the action that corresponds to the specific real-time state information of the appliance. A control policy is said to be optimal if it minimizes the expected cost function of the considered problem.} for $\{s_{i,k}\}$ that, based on the customer's specific real-time requests, can schedule the appliances in real time satisfying the real-time scheduling constraints \eqref{eq:HEMS c} and \eqref{eq:HEMS d} while minimizing the expected cost in \eqref{eq:HEMS a}. 
%
For the HEM design problem in \eqref{eq:HEMS}, however, it is intrinsically too difficult to obtain such optimal policy in its current form.} In Section \ref{subsec: MDP for HEM}, we will show that an optimal control policy of \eqref{eq:HEMS} can be efficiently obtained by reformulating \eqref{eq:HEMS} as a Markov decision process \cite{Shapiro2007}.
%

\vspace{-0.2cm}
\subsection{Proposed CoHEM Design}\label{subsec: proposed cohems}
\vspace{-0.0cm}

Dynamic prices $\{\pi(t)\}_{t=1}^T$ are designed by the retailer so that the customers would move their load to use cheaper electricity and mitigate congestion in the grid. As mentioned in the introduction, due to the rebound peaks, the aggregate power load from multiple HEM-based customers may not necessarily follow the bulk power purchase of the retailer, and the resultant real-time power imbalance will increase the cost of the retailer in the wholesale real-time balancing market.
{Specifically, in addition to purchasing electricity in the day-ahead and hour-ahead markets, the retailer has to purchase additional amount of energy in the real-time balancing market or pay the grid for absorbing the excessive energy that cannot be consumed.
The profit of the retailer is roughly
\begin{align}\label{eq: profit}
   \text{Profit}= B_{\rm c} - \text{Cost}_{RT} -\text{Cost}_{DHA}
\end{align} where $B_{\rm c}$ represents the total money paid by the customers for their electricity usage, $\text{Cost}_{DHA}$ denotes the cost already paid by the retailer in the day-ahead and hour-ahead markets, and $\text{Cost}_{RT}$ denotes the cost in the real-time market. As discussed, the selfish HEM systems will potentially increase $\text{Cost}_{RT}$. Hence it is desirable to coordinate the customers to reduce $\text{Cost}_{RT}$\footnote{The reduction in $\text{Cost}_{RT}$
implies that
the retailer can make more profits, and in turn the retailer may consider reducing the electricity price of the customers. Hence the CoHEM program is potentially economically beneficial to the cooperative customers from a long term perspective.}.
}

Let $\pi_{\rm p}(t)$ be the price for buying energy from the real-time market and $\pi_{\rm s}(t)$ be the price for absorbing extra energy (if $\pi_{\rm s}(t)\leq 0$ then it implies that the retailer may sell back the extra energy). Let $P(t)$ be the bulk power already purchased in the day-ahead and hour-ahead markets ($P(t)$ may also contain powers generated by local renewable sources, if applicable). Moreover, assume that there are a total of $H$ residences/customers, each contributing the power load $L_{\rm total}^{(h)}(t)$ [see \eqref{eq:total load}] to the grid.
The total real-time market cost of the retailer $\text{Cost}_{RT}$ is given by\footnote{The real-time cost can be extended to general convex functions; see Section III-C.}

{\small\begin{align}\label{eq:deviation cost}
 \!\!\!\!\text{Cost}_{RT}=&\sum_{t=1}^T \left[ \pi_{\rm s}(t)\left(P(t)-\sum_{h=1}^H L_{\rm total}^{(h)}(t)\right)^+ \notag \right.\\
    &\left.~~~~~~~~+\pi_{\rm p}(t)\left(\sum_{h=1}^H L_{\rm total}^{(h)}(t)-P(t)\right)^+ \right],
\end{align}}\hspace{-0.16cm} where $(x)^+=\max\{x,0\}$.
As incentives for the customers to participate in the proposed CoHEM architecture, we propose that 1) the retailer will charge the customers the same amount of money as that optimized by their individual (selfish) HEM system [i.e., \eqref{eq:HEMS}]; 2) the CoHEM will maintain the same scheduling deadline constraints specified by each of the customers. In summary, the CoHEM customers would neither have any financial loss nor would lose any degree of comfort.
{The retailer, which presumably will have infrastructure cost to cover, seeks to make a profit by minimizing $\text{Cost}_{RT}$  under the strict deadline constraints. Hence, the actual degree of flexibility of the community customers affects the retailer's profit.
}

The proposed CoHEM design problem can be formulated as
\vspace{-0.3cm}

{\small \begin{subequations}\label{eq:COHEMS}
\begin{align}
\!\!\!  \min_{\substack{s_{i,1}^{(h)},s_{i,2}^{(h)},\ldots\\ \forall~i,h}}~&{\small \sum_{t=1}^T \E\left[ \pi_{\rm s}(t)\left(P(t)-\sum_{h=1}^H L_{\rm total}^{(h)}(t)\right)^+ \notag \right.}\\
    &{\small \left.~~~~~~~+\pi_{\rm p}(t)\left(\sum_{h=1}^H L_{\rm total}^{(h)}(t)-P(t)\right)^+ \right]} \label{eq:COHEMS a} \\
  \text{s.t.}~& L_{\rm total}^{(h)}(t)=U^{(h)}(t) + \sum_{i=1}^ND_i^{(h)}(t),~\forall h,t,\label{eq:COHEMS b}\\
  & \text{constraints~in~}\eqref{eq:HEMS b},\eqref{eq:HEMS c},\eqref{eq:HEMS d}~\forall i,k,h,
\label{eq:COHEMS e}
\end{align}
\end{subequations}}where
we use superscript $(h)$ to denote the $h$th residence/customer; for example, $t_{i,k}^{(h)}$ and $s_{i,k}^{(h)}$ represent the request arrival and task operating times of appliance $i$ in the $h$th residence. 
{Similar to the selfish HEM design problem \eqref{eq:HEMS}, for the CoHEM problem \eqref{eq:COHEMS}, we aim to find an optimal control policy for $\{s_{i,k}^{(h)}\}$ for all $k,i$ and $h$, in order to schedule the appliances satisfying the real-time needs of the customers while minimizing the expected real-time market cost in \eqref{eq:COHEMS a}.} Note that problem \eqref{eq:COHEMS} is subject to the same scheduling constraints as \eqref{eq:HEMS} for each residence, meaning that the degrees of comfort of customers are preserved in the CoHEM architecture.
{However, different from \eqref{eq:HEMS} which is solved by each individual residence independently, the CoHEM problem \eqref{eq:COHEMS} and its associated scheduling policy have to be jointly optimized across all the residences in order to minimize the real-time market cost of the retailer.}

\vspace{-0.1cm}
\section{Proposed Solutions for HEM and CoHEM}\label{sec: methods}
\vspace{-0.0cm}
In this section, we present the detailed methods for solving the HEM design problem in \eqref{eq:HEMS} and handling the proposed CoHEM design problem in \eqref{eq:COHEMS}. In Section \ref{subsec: MDP for HEM},
we show that the HEM design problem \eqref{eq:HEMS} can be efficiently solved by reformulating it as a Markov decision process (MDP). {Regarding the CoHEM design problem \eqref{eq:COHEMS}, as discussed in the introduction section, our interest lies in a decentralized scheduling algorithm. To this end, we present in Section \ref{subsec: stochastic dual decomp} a suboptimal but efficient decentralized algorithm for handling \eqref{eq:COHEMS}, based on proper problem approximation and Lagrange dual optimization techniques \cite{Boyddecomposition}. It will be shown that the MDP method presented in Section \ref{subsec: MDP for HEM} can be conveniently employed by the proposed decentralized algorithm for computing the load scheduling policy of each individual residence.}
Finally, two interesting extensions of the proposed CoHEM architecture are presented in in Section \ref{subsection extensions}.

\vspace{-0.0cm}
\subsection{Solving HEM Problem \eqref{eq:HEMS} by MDP}\label{subsec: MDP for HEM}
\vspace{0.00cm}

\begin{figure*}[t!] \centering
   \resizebox{0.8\textwidth}{!}{
     \includegraphics{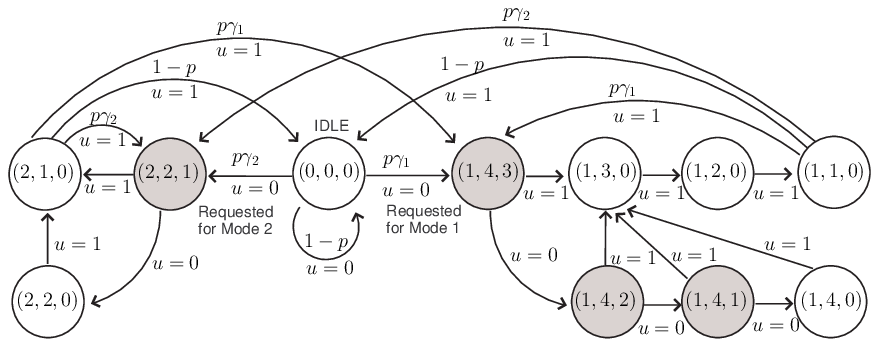}}\vspace{-0.0cm}
\caption{Illustration of the Markov process for modeling a 2-mode deferrable appliance with
$M_i=2$, $G_{i,1}=4$, $\zeta_{i,1}=3$, $G_{i,2}=2$, $\zeta_{i,2}=1$, $p_{i,1}(t)=p\gamma_1$ and $p_{i,2}(t)=p\gamma_2$. The triplets in the circles stand for the state $(S,W,Q)$ of the appliance.}
\label{fig:markov}\vspace{-0.3cm}
\end{figure*}

{In the subsection, we show how \eqref{eq:HEMS} can be efficiently solved by an MDP formulation. 
}
Note that, for the HEM design problem \eqref{eq:HEMS}, we can focus on optimizing one appliance, say appliance $i$, as follows
\begin{subequations}\label{eq:HEMS ind}
\begin{align}
  \min_{s_{i,1},s_{i,2},\ldots}~&\sum_{t=1}^T \E\left\{\pi(t) \left( \sum_{k=1}^{\infty} g_{i,\theta_i(t_{i,k})}(t- s_{i,k})\right)\right\} \label{eq:HEMS ind a} \\
  \text{s.t.}~& t_{i,k} \leq s_{i,k} \leq  t_{i,k}+\zeta_{i,\theta_i(t_{i,k})}~\forall k, \\
  & s_{i,k} \leq \max\{T-G_{i,\theta_i(t_{i,k})},t_{i,k}\}~ \forall k. \label{eq:HEMS ind d}
\end{align}
\end{subequations}
According to the appliance model described in Section \ref{subsec: load model}, once the customer requests to activate an appliance, the appliance is either operated immediately or it waits in a queue to be turned on later by the scheduler. We use the three variables $S_i(t) \in \{0,1,\ldots,M_i\}$,
$W_i(t) \in \{0,\ldots,\max_{m=1,\ldots,M}G_{i,m}\}$ and $Q_i(t)\in \{0,1,\ldots,\max_{m=1,\ldots,M}\zeta_{i,m}\}$ to denote the operation mode, the
remaining job length and the remaining maximum delay time for appliance $i$ at time $t$. Moreover, we use $u_i(t)\in \{0,1\}$ as a control variable that switches on and off the appliance at time $t$.

When all $S_i(t),W_i(t)$ and $Q_i(t)$ are zero (i.e., $(S_i(t),W_i(t),Q_i(t))=(0,0,0)$), the appliance is idle, and consequently $u_i(t)=0$. Then it is with probability $p_i(t+1)\gamma_{i,m}(t+1)$ that the customer will request to activate appliance $i$ at time $t+1$ and operate it in mode $m$. If appliance $i$ is called by the customer to operate in mode $m$, then we have $(S_i(t+1),W_i(t+1),Q_i(t+1))=(m,G_{i,m},\zeta_{i,m})$, and the HEM system has to decide whether to activate the appliance or to queue the task by deciding $u_i(t+1)=1$ or $u_i(t+1)=0$. If the controller chooses to activate the appliance, then the appliance has to work for $G_{i,m}$ time slots consecutively, following its load profile $g_{i,m}(1), \ldots, g_{i,m}(G_{i,m})$ (i.e., $u_i(t+1)=u_i(t+2)=\cdots=u_i(t+G_{i,m})=1$). After that, it is with probability $p_i(t+G_i+1)\gamma_{i,m'}(t+G_i+1)$ that the appliance is called (for mode $m'$) and goes back to state $(m',G_{i,m'},\zeta_{i,m'})$; otherwise (with probability $1-p_i(t+G_i+1)$) it goes back to state $(0,0,0)$ (idle).

If, at state $(m,G_{i,m},\zeta_{i,m})$, the controller chooses to delay the appliance ($u_i(t+1)=0$), then the state moves to $(S_i(t+2),W_i(t+2),Q_i(t+2))=(m,G_{i,m},\zeta_{i,m}-1)$ in the next time slot, and the HEM system needs to decide if $u_i(t+2)$ should be one or zero. The decision process is repeated until the HEM system chooses to turn on the appliance or until the maximum delay time is reached. In both cases, the state moves to $(m,G_{i,m}-1,0)$. An example of a 2-mode appliance is illustrated in Figure \ref{fig:markov}.

Let $\Xb_i(t)=(S_i(t),W_i(t),Q_i(t))$ be a state vector, {and
$\mathcal{\Xb}_i$ be the set that contains all possible $\Xb_i(t)$'s.}
Using the above Markov process model, the HEM design problem \eqref{eq:HEMS ind} can be rewritten as the following MDP:
\begin{subequations}\label{eq:HEMS Markov}
\begin{align}
\!\!\!\!\!\!  {\min_{\substack{\mu_{i,t}(\Xb),~\\\forall \Xb\in\mathcal{X}_i,\\t=1,\ldots,T }}}~&\sum_{t=1}^T \E\left\{\pi(t) \left(u_i(t)g_{i,S_i(t)}(G_{i,S_i(t)}-W_i(t)+1)\right)\right\} \notag \\
  \text{s.t.}~
  & \mu_{i,t}(\Xb_i(t))=u_i(t),  \\
  & \Xb_i(t+1)=\begin{bmatrix}
  \mathcal{S}_i(\Xb_i(t),\theta_i(t+1)) \\
  \mathcal{W}_i(\Xb_i(t),u_i(t),\theta_i(t+1)) \\
  \mathcal{Q}_i(\Xb_i(t),u_i(t),\theta_i(t+1))
  \end{bmatrix} \notag \\
  &~~~~~~~~~~~~~~~~~~~~~~~~~~~~~~~~~~~\forall t=1,\ldots,T, \label{eq:HEMS Markov b}\\
  & \mu_{i,t}(S_i(t),G_{i,m},Q_i(t))\!=\!1~ \forall~t\!\geq\! T-G_{i,S_i(t)},  \label{eq:HEMS Markov c}
 \end{align}
 \end{subequations}
where $\theta_i(t+1)\in \{0,1,\ldots,M_i\}$ is a random variable with $\Prob(\theta_i(t+1)=m)=p_i(t+1)\gamma_{i,m}(t+1)$ and $\Prob(\theta_i(t+1)=0)=1-p_i(t+1)$,
indicating the operation mode of the appliance at time $t+1$ provided that it is requested. Moreover, $\mathcal{S}_i(\cdot)$, $\mathcal{W}_i(\cdot)$ and $\mathcal{Q}_i(\cdot)$ are the state transition functions given by (see Figure \ref{fig:markov} as an example)
\begin{align}
  &\mathcal{S}_i(\Xb_i(t),\theta_i(t+1))=\left\{\!\!\!\!\!\begin{array}{ll}
    &\theta_i(t+1)~\text{if}~S_i(t)=0, \notag \\
    &S_i(t)~~~~~~\text{otherwise},\notag
  \end{array}\right. \notag 
  \end{align}
  \begin{align}
  &\mathcal{W}_i(\Xb_i(t),u_i(t),\theta_i(t+1))\notag\\
  &=\left\{\!\!\!\!\!\begin{array}{ll}
    &W_i(t)-1~\text{if}~S_i(t)=m,2 \leq W_i(t) \leq G_{i,m},u_i(t)=1, \notag \\
    &W_i(t)~~~~~~\text{if}~ W_i(t) \geq 1,u_i(t)=0, \notag \\
    &G_{i,m}~~~~~~~\text{if}~0 \leq W_i(t)\leq 1~\text{and}~\theta_i(t+1)=m, \notag\\
    &0~~~~~~~~~~~~\text{if}~0 \leq W_i(t)\leq 1~\text{and}~\theta_i(t+1)=0,\notag
  \end{array}\right. \notag \\
  &\mathcal{Q}_i(\Xb_i(t),u_i(t),\theta_i(t+1))\notag\\
  &=\left\{\!\!\!\!\!\!\!\!\!\begin{array}{ll}
    &Q_i(t)-1~~\text{if}~S_i(t)=m,W_i(t)= G_{i,m},u_i(t)=0, \notag \\
    &\zeta_{i,m}~~~~~~~~~\text{if}~0\leq W_i(t)\leq 1,~\theta_i(t+1)=m, \notag\\
    &0~~~~~~~~~~~~\text{if}~S_i(t)=m,2 \leq W_i(t)\leq G_{i,m},u_i(t)=1,  \notag\\
    &~~~~~~~~~~\text{or}~\text{if}~W_i(t)=1~\text{and}~\theta_i(t+1)=0,
    \notag
  \end{array}\right.
\end{align} and \eqref{eq:HEMS Markov c} is due to \eqref{eq:HEMS ind d}, which enforces the requested appliance to turn on if its task is still queued for $t\geq T-G_{i,m}$.
{In \eqref{eq:HEMS Markov}, the optimal control policy for $u_i(t)$, denoted by $\mu_{i,t}(\Xb_i(t))$, is a function of $\Xb_i(t)$, for all $\Xb_i(t) \in \mathcal{X}_i$
and for all $t=1,\ldots,T$. Once the control policy $\mu_{i,t}(\cdot)$, $t=1,\ldots,T$, is obtained, the controller can schedule the appliance in real time, following the policy given the real-time state status $\Xb_i(t)$ of the appliance.}

By dynamic programming (DP) \cite{BK:Bersekas07}, $\mu_{i,t}(\Xb_i(t))$ can be obtained by considering the following backward recursive equations
\begin{align}\label{eq: HEM Bellman}
  \!\!J_t(\Xb_i(t))\!=\!\!\min_{u_i(t)\in \{0,1\}}~&\pi(t)u_i(t)g_{i,m}(G_{i,m}-W_i(t)+1)+\notag\\
  &~~~~~~~\E_{\theta_i(t+1)}\{J_{t+1}(\Xb_i(t+1))\} \notag\\
  \text{s.t.}~&S_i(t)=m, \notag \\
  &\text{constraints~in~}\eqref{eq:HEMS Markov b},\eqref{eq:HEMS Markov c},
\end{align}
for all possible $\Xb_i(t)$, $t=1,\ldots,T-1$. Specifically, $\mu_{i,t}(\Xb_i(t))$ is given by the optimal $u_i(t)$ of \eqref{eq: HEM Bellman}.
The value of $J_t(\Xb_i(t))$ is known as the cost-to-go function \cite{BK:Bersekas07}.
One can show that \eqref{eq: HEM Bellman} can be classified into the following four cases (see Figure \ref{fig:markov} as an example):

\noindent \underline{\emph{Case 1} ($S_i(t)=0$,$W_i(t)=0$, $Q_i(t)=0$):} $\mu_{i,t}(0,0,0)=0$, and
\begin{align}\notag
  J_t(0,0,0) =&(1-p_i(t+1))J_{t+1}(0,0,0) + \notag \\
  &~~~p_i(t+1)\sum_{m=1}^{M_i} \gamma_{i,m}J_{t+1}(m,G_{i,m},\zeta_{i,m}).
\end{align}

\noindent \underline{\emph{Case 2} ($S_i(t)=m$, $W_i(t)=G_{i,m}$, $1 \leq Q_i(t) \leq \zeta_{i,m}$):}  For $t\geq T-G_{i,m}$, $\mu_{i,t}(m,G_{i,m},Q_i(t))=1$ according to \eqref{eq:HEMS Markov c}; otherwise $\mu_{i,t}(m,G_{i,m},Q_i(t))$ is given by the optimal solution of the following problem
\begin{align}
  &J_t(m,G_{i,m},Q_i(t)) \notag \\
  &=\min_{u_i(t)\in \{0,1\}}~\pi(t)u_i(t)g_{i,m}(1)+\notag\\
  &~~~~~~~~~~~~~~~~~~~~~~~~~~~J_{t+1}(m,W_i(t+1),Q_i(t+1)) \notag\\
  &~~~~~~~~~~~~\text{s.t.}~W_i(t+1)=G_{i,m}-u_i(t), \notag\\
  &~~~~~~~~~~~~~~~~~Q_i(t+1)=\left\{\!\!\!\!\!\!\!\!\!\begin{array}{ll}
    &0,~~\text{if}~u_i(t)=1, \notag \\
    &Q_i(t)-1,~~\text{if}~u_i(t)=0.
  \end{array}\right.
\end{align} 

\noindent \underline{\emph{Case 3} ($S_i(t)=m$, $2 \leq W_i(t)\leq G_{i,m}$, $Q_i(t)=0$):} $\mu_{i,t}(m,W_i(t),0)=1$, and
\begin{align}\notag
  J_t(m,W_i(t),0) =&\pi(t)g_{i,m}(G_{i,m}-W_i(t)+1)+ \notag \\
  &~~~~~~~~~~~~~~\!J_{t+1}(m,W_i(t)-1,0). \notag
\end{align}

\noindent \underline{\emph{Case 4} ($S_i(t)=m$, $W_i(t)=1$, $Q_i(t)=0$):} $\mu_{i,t}(m,1,0)=1$, and
\begin{align}\notag
\!\!\!\!\!  J_t(m,1,0) =&\pi(t)g_{i,m}(G_{i,m}) + (1-p_i(t+1))J_{t+1}(0,0,0)\notag \\
  &~~~~+p_i(t+1)\sum_{m=1}^{M_i} \gamma_{i,m}J_{t+1}(m,G_{i,m},\zeta_{i,m}).\notag
\end{align}

It can be seen that, only for Case 2 (e.g., the gray states shown in Figure \ref{fig:markov}), one is required to compute the optimal control policy $\mu_{i,t}(\Xb_i(t))$. Therefore, at each stage of the backward computation, one needs only compute and store $\mu_{i,t}(\Xb_i(t))$ for $\sum_{m=1}^{M_i}\zeta_{i,m}$ possible states. Since both $M_i$ and $\zeta_{i,m}$ are usually small numbers, the backward computation can be implemented very efficiently.
By applying the MDP approach to each of the appliances, we obtain the optimal appliance control policy for the HEM design problem \eqref{eq:HEMS}.
\begin{Remark}{\rm
In the Markov model presented above, the appliance has multiple modes which are decided by the customer. Sometimes, each mode may have multiple power consumption alternates that all can fulfill the same task. In that case, the controller has a further degree of freedom to choose the one that is best out of all the available consumption profiles in minimizing the customer's cost. This appliance model can be easily extended from the current Markov model described in this subsection. Specifically, suppose that, for an appliance, each mode has 2 alternate power profiles to choose from, e.g, $g_{i,m,1}(t)$ and $g_{i,m,2}(t)$. The control variable $u_i(t)$ then has three possible values -- 0 means WAIT (or remain IDLE if not requested), 1 means ON following profile 1, and 2 means ON following profile 2. The state vector for this case is given by $\Xb_i(t)=(S_i(t),R_i(t),W_i(t),Q_i(t))$ where $R_i(t)\in \{0,1,2\}$. An example for
$M_i=1$ (single mode), $G_{i,1}=3$, $\zeta_{i,1}=2$ and $p_{i,1}(t)=p$ is illustrated in Fig. 2. The associated optimal control policy for the HEM design problem \eqref{eq:HEMS} can be derived in a similar fashion as described in this subsection.
}
\end{Remark}
\vspace{-0.2cm}

\begin{figure}[t!] \centering
   \resizebox{0.5\textwidth}{!}{
     \includegraphics{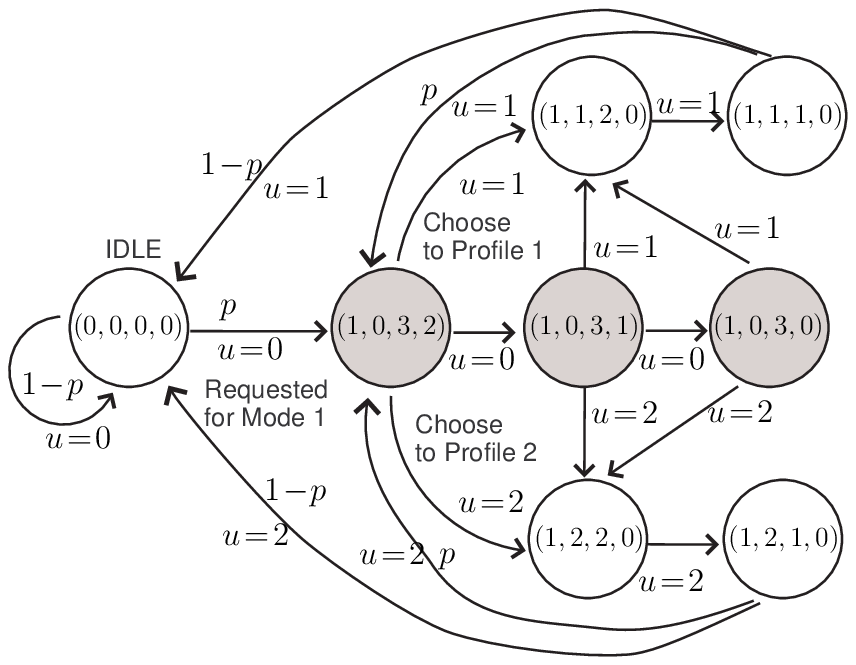}}\vspace{-0.0cm}
\caption{Illustration of the Markov process for modeling a single-mode deferrable appliance $(M_i=1)$ with
two alternate power consumption profiles, and $G_{i,1}=3$, $\zeta_{i,1}=2$ and $p_{i,1}(t)=p$. The quadruples in the circles stand for the state $(S,R,W,Q)$ of the appliance; see Remark 1.}
\label{fig:markov_2s}\vspace{-0.4cm}
\end{figure}

\vspace{-0.4cm}
\subsection{Decentralized Stochastic Optimization for CoHEM}\label{subsec: stochastic dual decomp}

While the Markov process model presented in the previous subsection can also be used for the CoHEM problem \eqref{eq:COHEMS}, the resultant MDP problem would involve a large number of states (which increases exponentially with $H$ and $N$) {and thus the solution quickly becomes computationally unaffordable.}
A simple approximate DP approach to overcoming the curse of dimensionality issue is the model predictive certainty equivalent control (CEC) method \cite{BK:Bersekas07}. In this method, one searches for the optimal control in a forward manner and apply the control at each time that would be optimal if the uncertain quantities, i.e., $t_{i,k}^{(h)}$ and $\theta_{i}^{(h)}(t_{i,k})$, were fixed at some typical values, e.g., the mean values.
The advantage of this method is that at each time $t$, one only deals with a deterministic optimization problem; the associated solution can be directly applied to control the appliances in real time.
In \cite{ChangPES12}, we have proposed a decentralized CoHEM algorithm (i.e., \cite[Algorithm 1]{ChangPES12}) based on the model predictive CEC method. While the algorithm in \cite{ChangPES12} can be implemented in a decentralized fashion (i.e., \cite[Algorithm 2]{ChangPES12}), the controller has to repeat the optimization until the end of the horizon, from time 1 to time $T-1$. Moreover, since the decentralized optimization algorithm involves multiple iterations, each of which requires the customers to exchange messages with their neighbors in real time.
As a result, the model predictive CEC algorithm in \cite{ChangPES12} may cause a considerable communication overhead in the CoHEM network.
In this subsection, we present a decentralized stochastic optimization method which can greatly alleviate the communication overhead.

Unlike the model predictive CEC, the idea here is to maintain the stochastic nature of the CoHEM problem \eqref{eq:COHEMS}, using a suboptimal formulation as follows
\begin{subequations}\label{eq:COHEMS lowerbound}
\begin{align}
\!\!\!\!\!\!  \min_{\substack{s_{i,1}^{(h)},s_{i,2}^{(h)},\ldots\\ \forall~i,h}}~&\sum_{t=1}^T \bigg\{ \pi_{\rm s}(t)\left(\hat{P}(t)-\sum_{h=1}^H \sum_{i=1}^N\E[D_i^{(h)}(t)]\right)^+ \notag \bigg. \\
    &~\bigg.+\pi_{\rm p}(t)\left(\sum_{h=1}^H \sum_{i=1}^N\E[D_i^{(h)}(t)]-\hat{P}(t)\right)^+\bigg\}  \label{eq:COHEMS lowerbound a} \\
  \text{s.t.}~&
   D_i^{(h)}(t)=\sum_{k=1}^{\infty} g_i^{(h)}(t- s_{i,k}^{(h)})~\forall i,t,h, \label{eq:COHEMS lowerbound c}\\
  & t_{i,k}^{(h)} \leq s_{i,k}^{(h)} \leq  t_{i,k}^{(h)}+\zeta_{i,\theta_i^{(h)}(t_{i,k}^{(h)})}^{(h)}~\forall i,k,h, \label{eq:COHEMS lowerbound d}\\
  & s_{i,k}^{(h)} \leq \max\{T-G_{i,\theta_i^{(h)}(t_{i,k}^{(h)})}^{(h)},t_{i,k}^{(h)}\}~ \forall i,k,h, \label{eq:COHEMS lowerbound e}
  \end{align}
\end{subequations}
where $\hat{P}(t)=P(t)-\sum_{h=1}^H U^{(h)}(t)$. {It can be seen that the objective function \eqref{eq:COHEMS lowerbound a} is instead the deviation between the expected aggregate load and the power supply $\hat{P}(t)$, which
is a lower bound of that in \eqref{eq:COHEMS a}.}
Note that \eqref{eq:COHEMS lowerbound} is still a (multi-stage) stochastic optimization problem. 

Define
\begin{align}
  z(t)=\left(\sum_{h=1}^H \sum_{i=1}^N\E[D_i^{(h)}(t)]-\hat{P}(t)\right)^+~\forall t,
\end{align}
and note that
{\small\begin{align}
 &\left(\hat{P}(t)-\sum_{h=1}^H \sum_{i=1}^N\E[D_i^{(h)}(t)]\right)^+ \notag \\
 &~~~~~~~~~~~~~~~~~~=
 z(t)-\left(\sum_{h=1}^H \sum_{i=1}^N\E[D_i^{(h)}(t)]-\hat{P}(t)\right).
\end{align}}
Substituting the above equations into \eqref{eq:COHEMS lowerbound} gives rise to
\begin{subequations}\label{eq:COHEMS lowerbound1}
\begin{align}
\!\!\!  \min_{\substack{s_{i,1}^{(h)},s_{i,2}^{(h)},\ldots\\ \forall~i,h}}~&\sum_{t=1}^T
(\pi_{\rm s}(t)+\pi_{\rm p}(t))z(t)\notag \\
&-\sum_{t=1}^T\pi_{\rm s}(t)\left(\sum_{h=1}^H \sum_{i=1}^N\E[D_i^{(h)}(t)]-\hat{P}(t)\right) \label{eq:COHEMS lowerbound1 a} \\
  \text{s.t.}~&
   z(t)=\!\! \left(\sum_{h=1}^H \sum_{i=1}^N \E[D_i^{(h)}(t)]-\hat{P}(t)\right)^+\!\!\forall t, \label{eq:COHEMS lowerbound1 c}\\
  & \text{constraints in } \eqref{eq:COHEMS lowerbound c},\eqref{eq:COHEMS lowerbound d},\eqref{eq:COHEMS lowerbound e}. \label{eq:COHEMS lowerbound1 d}
\end{align}
\end{subequations}
Assuming the usual case of $\pi_{\rm s}(t)+\pi_{\rm p}(t)\geq 0$, 
we can further rewrite \eqref{eq:COHEMS lowerbound1} as
\begin{subequations}\label{eq:COHEMS lowerbound2}
\begin{align}
\!\!\!  \min_{\substack{s_{i,1}^{(h)},s_{i,2}^{(h)},\ldots\\ \forall~i,h}}~&\sum_{t=1}^T
(\pi_{\rm s}(t)+\pi_{\rm p}(t))z(t)\notag \\
&-\sum_{t=1}^T\pi_{\rm s}(t)\left(\sum_{h=1}^H \sum_{i=1}^N\E[D_i^{(h)}(t)]-\hat{P}(t)\right) \label{eq:COHEMS lowerbound2 a} \\
  \text{s.t.}~&
   z(t)\geq \sum_{h=1}^H \sum_{i=1}^N \E[D_i^{(h)}(t)]-\hat{P}(t)~\forall~t,
    \label{eq:COHEMS lowerbound2 c}\\
    &z(t) \geq 0 ~\forall t, \label{eq:COHEMS lowerbound2 c2}\\
  & \text{constraints in } \eqref{eq:COHEMS lowerbound c},\eqref{eq:COHEMS lowerbound d},\eqref{eq:COHEMS lowerbound e}. \label{eq:COHEMS lowerbound2 d}
\end{align}
\end{subequations}

{The ingredient of our decentralized stochastic optimization method is to consider the Lagrange dual \cite{BK:Boyd04} of the stochastic problem \eqref{eq:COHEMS lowerbound2}, analogous to the methods adopted in \cite{Williams2007,Furmston2011} for other applications.} Let $\lambda(t)\geq 0$ and $\eta(t)\geq 0$ be the dual variables associated with each of the constraints in \eqref{eq:COHEMS lowerbound2 c} and \eqref{eq:COHEMS lowerbound2 c2}. The dual function of \eqref{eq:COHEMS lowerbound2} can be shown to be
\begin{align*}
\left\{\!\!\!\!\!
        \begin{array}{ll}
        &{\displaystyle \sum_{h=1}^H \Phi^{(h)}(\lambdab),
        }~~~~\text{if}~\pi_{\rm p}(t)+\pi_{\rm s}(t)-\lambda(t)=\eta(t)~\forall t, \\
        &-\infty,~~~~~~~~~~~~~\text{elsewhere},
        \end{array}
        \right.
\end{align*}
where $\lambdab=[\lambda(1),\ldots,\lambda(T)]^T$ and
{ \small\begin{align}\label{eq: stochastic subprob}
  &\Phi^{(h)}(\lambdab)=\notag \\
  &\min_{s_{i,1}^{(h)},s_{i,2}^{(h)},\ldots}
  \sum_{t=1}^T\E\!\left[\!(\lambda(t)-\pi_{\rm s}(t)) \left(\sum_{i=1}^ND_i^{(h)}(t)+U^{(h)}(t)\!-\! \frac{P(t)}{H}\right)\right] \notag \\
  &~~~~~~~~~~\text{s.t.}~\text{constraints in } \eqref{eq:COHEMS lowerbound c},\eqref{eq:COHEMS lowerbound d},\eqref{eq:COHEMS lowerbound e}.
\end{align}}
{The dual problem of \eqref{eq:COHEMS lowerbound2} is thus given by
\begin{subequations}\label{eq:COHEMS stochastic dual0}
\begin{align}
\max_{\substack{\lambda(t), \\t=1,\ldots,T}}~
&\sum_{h=1}^H \Phi^{(h)}(\lambdab)
 \\
\text{s.t.}&~\lambda(t) \geq 0~\forall~t=1,\ldots,T, \label{eq:COHEMS stochastic dual0 C1}\\
           &~\eta(t) \geq 0,~\forall~t=1,\ldots,T, \\
           &~\pi_{\rm p}(t)+\pi_{\rm s}(t)-\lambda(t)=\eta(t)~\forall t=1,\ldots,T.
           \label{eq:COHEMS stochastic dual0 C3}
\end{align}
\end{subequations}
One can see from the above equation that $\eta(t)$ is in fact a dummy variable since it does not appear in the objective function. By combining \eqref{eq:COHEMS stochastic dual0 C1} to \eqref{eq:COHEMS stochastic dual0 C3}, we then obtain }
\begin{align}\label{eq:COHEMS stochastic dual}
\max_{\substack{\lambda(t), \\t=1,\ldots,T}}~
&\sum_{h=1}^H \Phi^{(h)}(\lambdab)
 \notag \\
\text{s.t.}&~0\leq \lambda(t) \leq \pi_{\rm s}(t)+\pi_{\rm p}(t)~\forall~t=1,\ldots,T.
\end{align}
{The dual optimization method for \eqref{eq:COHEMS lowerbound2} is to iteratively solve the inner minimization problems in \eqref{eq: stochastic subprob} and the outer maximization part in \eqref{eq:COHEMS stochastic dual} \cite{Boyddecomposition}.

{\bf Distributed inner primal minimization:}
Let $\lambda(t;n)$ denote the dual variable $\lambda(t)$ obtained at iteration $n$. Given
$\lambda(t;n),t=1,\ldots, T,$ the algorithm solves the inner minimization problems in \eqref{eq: stochastic subprob} for all $h=1,\ldots,H$. Note that the objective function of \eqref{eq:COHEMS stochastic dual} is a summation of $\Phi^{(h)}(\lambdab),h=1,\ldots, H,$ which is decomposable. Thus the inner minimization step can be carried out in a fully distributed fashion where each residence $h$ solves the corresponding subproblem \eqref{eq: stochastic subprob} independently.

It is important to observe that subproblem \eqref{eq: stochastic subprob} has exactly the same formulation as the selfish HEM design problem \eqref{eq:HEMS}, except that, in \eqref{eq: stochastic subprob}, the $h$th residence is given a ``pseudo price" $\lambda(t;n)-\pi_{\rm s}(t),t=1,\ldots, T$.} Therefore, the MDP method presented in Section \ref{subsec: MDP for HEM} can be directly used to efficiently solve \eqref{eq: stochastic subprob} and obtain the optimal appliance control policy of \eqref{eq: stochastic subprob} for each residence $h$.
{Let us denote $\{\mu^{(h)}_{i,t}(\cdot;n)\}_{t=1}^T$, $i=1,\ldots,N$, (see \eqref{eq:HEMS Markov}) as the optimal appliance control policy for  \eqref{eq: stochastic subprob} obtained by residence $h$ at iteration $n$. Moreover, let $\E\{D_i^{(h)}(t;n)\}$, $i=1,\ldots,N$, $t=1,\ldots,T$, be the corresponding expected scheduled loads for residence $h$.}

{\bf Distributed dual subgradient update:}
The dual variable $\lambda(t),$ $t=1,\ldots,T$, can be updated by the subgradient projection method \cite{Boydsubgradient}:
\begin{align}\label{eq:dual update2}
\!\!\!  \lambda(t;n+1)\!=&\mathcal{P}\left\{\!\!\lambda(t;n)\!+\!c_n\!\left(\sum_{h=1}^H \sum_{i=1}^N\E[D_i^{(h)}(t;n)]-\hat P(t)\!\!\right)\!
  \right\} \notag \\
  &~~~~~~~~~~~~~~~~~~~~~~~~~~~~~\forall t=1,\ldots,T,
\end{align}
where $c_n>0$ is the step size, and $\mathcal{P}(\cdot)$ denotes the operation of projection onto the set $[0, \pi_{\rm p}(t)+\pi_{\rm s}(t)]$.

In view of the fact that updating \eqref{eq:dual update2} requires the aggregate load $\sum_{h=1}^H \sum_{i=1}^N\E[D_i^{(h)}(t;n)]$ of all residences,
it usually requires a control center to coordinate the residences for the dual update. To perform the dual update \eqref{eq:dual update2} in a \emph{fully decentralized} fashion, we alternatively employ the consensus-subgradient method \cite{Johansson_CDC08}.
In this method, each residence $h$ maintains a set of local copies of $\lambda(t;n),$ $t=1,\ldots,T$, denoted by $\lambda^{(h)}(t;n)$, $t=1,\ldots,T$ on its own, and locally updates them according to the subgradient of $\Phi^{(h)}(\lambdab)$:
\begin{align}\label{eq:dual update3}
  &\nu^{(h)}(t;n)=\lambda^{(h)}(t;n)+ \notag \\
   &~~~~~~~~~~~\!c_n\!\left( \sum_{i=1}^N\E[D_i^{(h)}(t;n)]+\E[U^{(h)}(t)]- \frac{P(t)}{H}\!\!\right)\!
\end{align}
for $t=1,\ldots,T$. Note that, in \eqref{eq:dual update3}, only the local information about residence $h$ is used, in addition to the bulk power purchase $P(t)$ {broadcasted by the retailer.} Since the desired subgradient update \eqref{eq:dual update2} is the average of \eqref{eq:dual update3} over all $h=1,\ldots,H$, the second step in the consensus-subgradient method \cite{Johansson_CDC08} is that each residence $h$ exchanges with its connecting neighbors about $\nu^{(h)}(t;n)$ so as to achieve a consensus on $\lambda(t;n+1)$. More precisely, residence $h$ obtains $\lambda^{(h)}(t;n+1)$ by
\begin{align}\label{eq:dual update5}
\!\!\!\!\lambda^{(h)}(t;n+1)=\mathcal{P}\left\{
    f^{\psi}(\nu^{(j)}(t;n), j\in \{h\}\cup\mathcal{N}_{h})\!
  \right\}
\end{align}
for $t=1,\ldots,T$, where $\mathcal{N}_{h}$ denotes the index set of the neighbors that can communicate with residence $h$,
$f(\cdot)$ is an averaging consensus function {(e.g., \cite{Johansson_CDC08}
$$
 f(\nu^{(j)}(t;n), j\in \{h\}\cup\mathcal{N}_{h})=\sum_{j\in \{h\}\cup\mathcal{N}_{h}}
 [\Wb]_{h,j} \nu^{(j)}(t;n),
$$
where $\Wb$ is an $H$ by $H$ mixing matrix),}
and the superscript $\psi$ indicates that the averaging consensus step is repeated for $\psi$ times. 

{The convergence properties of the consensus-subgradient method has been studied in \cite{Johansson_CDC08}. Roughly speaking, the dual iterates $\lambda^{(h)}(t;n+1)$ for all $h$ asymptotically converge to each other with a discrepancy no larger than $\epsilon_1>0$, and the corresponding dual objective value in \eqref{eq:COHEMS stochastic dual} also asymptotically approaches the optimal value with a deviation no more than $\epsilon_2>0$, where both $\epsilon_1$ and $\epsilon_2$ are small positive numbers and decrease with $\psi$. This implies that, if the number of consensus steps $\psi$ is sufficiently large, the consensus-subgradient method converges to the optimal solution of \eqref{eq:COHEMS stochastic dual}.  In computer simulations, we find that a small number of $\psi$ (e.g., $\psi \leq 15$) is sufficient for achieving good convergence performance.}

{\bf Monte Carlo method for estimating $\sum_{i=1}^N\E\{D_i^{(h)}(t;n)\}$:} {In order to perform the dual update \eqref{eq:dual update3}, each residence $h$ has to compute the expected scheduled load $\sum_{i=1}^N\E\{D_i^{(h)}(t;n)\}$,  $t=1,\ldots,T$, associated with the control policy $\{\mu^{(h)}_{i,t}(\cdot;n)\}_{t=1}^T$, $i=1,\ldots,N$. While it is difficult to obtain this expected power load analytically, one can estimate it through the Monte Carlo method \cite{Shapiro2007}.} In particular, the HEM unit in residence $h$ can repeatedly generate realizations of appliance requests $t_{i,1}^{(h)}$, $t_{i,2}^{(h)}$, \ldots (according to the customer's usage probabilities $p^{(h)}_i(t)$ and $\gamma^{(h)}_{i,m}(t)$), followed by applying the optimal control policy $\{\mu^{(h)}_{i,t}(\cdot;n)\}_{t=1}^T$, $i=1,\ldots,N,$ at iteration $n$; {this outputs a simulated scheduled load profile $\sum_{i=1}^ND_i^{(h)}(t)$, $t=1,\ldots,T$.
The HEM unit repeats this simulation multiple times, each of which outputs a scheduled load profile $\sum_{i=1}^ND_i^{(h)}(t)$, $t=1,\ldots,T$.
By averaging $\sum_{i=1}^ND_i^{(h)}(t)$, $t=1,\ldots,T$, over all the simulated realizations, residence $h$ can use this sample average as an estimate of $\sum_{i=1}^N\E\{D_i^{(h)}(t;n)\}$, $t=1,\ldots,T$.}
{Note that the computations mentioned above all can be implemented for each appliance and for each realization in a parallel manner.}

We summarize {the} decentralized stochastic optimization algorithm in Algorithm 1.
{Three remarks regarding Algorithm 1 are in order:}

\begin{algorithm}[h]
  \caption{{Decentralized stochastic optimization algorithm for \eqref{eq:COHEMS lowerbound}}}
\begin{algorithmic}[1]\label{alg:decentralized_modified}
  \STATE {\bf Input} an initial set of $\lambda^{(h)}(t;0)$, $t=1,\ldots,T$, at residence $h$, for all $h=1,\ldots,H$.
  \STATE Set $n=0$.
  \REPEAT
    \FOR{$h=1,\dots, H$}
      \STATE 1) {Given $\{\lambda^{(h)}(t;n)-\pi_{\rm s}(t)\}_{t=1}^T$, residence $h$
      solves \eqref{eq: stochastic subprob} by the MDP method in Section \ref{subsec: MDP for HEM} to obtain the optimal control policy of appliances $\{\mu^{(h)}_{i,t}(\cdot;n)\}_{t=1}^T,i=1,\ldots,N$.}
      \STATE 2) Residence $h$ {applies the instantaneous control policy} $\{\mu^{(h)}_{i,t}(\cdot;n)\}_{t=1}^T,i=1,\ldots,N$, and the Monte Carlo method to estimate the expected load
       $\sum_{i=1}^N\E\{D_i^{(h)}(t;n)\},t=1,\ldots,T.$

      \STATE 3) Residence $h$ obtains $\nu^{(h)}(t;n)$ by \eqref{eq:dual update3}.
      \STATE 4) Residence $h$ exchanges $\nu^{(h)}(t;n)$ with its connecting neighbors for updating
                $\lambda^{(h)}(t;n+1)$ by \eqref{eq:dual update5}. 
    \ENDFOR
   \STATE $n=n+1$
  \UNTIL the predefined stopping criterion is satisfied.
   \STATE All residences respectively apply {the running-averaged control polices in \eqref{eq:running average2}} for \emph{on-line} (real-time) scheduling.
\end{algorithmic}
\end{algorithm}

\vspace{-0.3cm}
\begin{Remark} {\rm We should emphasize that the above optimization method for \eqref{eq:COHEMS stochastic dual} can be done in an \emph{off-line} fashion since solving \eqref{eq:COHEMS stochastic dual} uses only the statistical information of customers (i.e., $p_i^{(h)}(t),\gamma_{i,m}^{(h)}(t)$).
As a result, we only need to perform the decentralized optimization \emph{once} for each look-ahead horizon $T$. 
The associated optimal control policy of appliances can then be applied to the real-time scheduling process. {Note that this is very different from the model predictive CEC method in \cite{ChangPES12}
where distributed optimization has to be carried out $T-1$ times. Therefore, the proposed Algorithm 1 has a much reduced computation and communication overheads for the CoHEM network. We should also mention that Algorithm 1 works well under the assumption that the retailer has reasonably accurate estimates of the real-time balancing prices $\{\pi_{\rm s}(t)\}_{t=1}^T$ and $\{\pi_{\rm p}(t)\}_{t=1}^T$.
}}
\end{Remark}

\begin{Remark} {\rm Since the proposed approach is based on the dual optimization \eqref{eq:COHEMS stochastic dual}, the primal control policy $\{\mu^{(h)}_{i,t}(\cdot;n)\}_{t=1}^T,i=1,\ldots,N$, obtained at each iteration $n$, may not converge as well as the dual variables $\{\lambda^{(h)}(t;n)\}_{t=1}^T$. In that case, following the same spirit as in \cite{Angelia2009,Larsson1999}, one can alternatively use the running averaged policy
\begin{align}\label{eq:running average2}
  {\rm round}\left\{\frac{1}{n}\sum_{\ell=1}^{n}\mu^{(h)}_{i,t}(\Xb;\ell)\right\}
  \in \{0,1\}~\forall \Xb\in \mathcal{X}_{i}^{(h)},t,i,h,
\end{align}
{(see Step 12 of Algorithm 1),} where the operator ${\rm round}(\cdot)$ rounds the averaged policy to its feasible region. We find through simulations that this running-averaged policy works well in practice.
}
\end{Remark}

\vspace{-0.5cm}
\begin{Remark} {\rm The proposed method is suboptimal compared to the original CoHEM design problem \eqref{eq:COHEMS} since it is optimizing the lower-bound problem in \eqref{eq:COHEMS lowerbound2}, and { \eqref{eq:COHEMS lowerbound2} and its Lagrange dual in \eqref{eq:COHEMS stochastic dual} have a non-zero duality gap in general \cite{Williams2007,Furmston2011}.} The suboptimiality can be measured as follows. Let us denote the optimal objective value of \eqref{eq:COHEMS} by $f^\star_p$, denote that of the lower-bound problem \eqref{eq:COHEMS lowerbound2} by $f_l^\star$, and denote that of the dual problem \eqref{eq:COHEMS stochastic dual} by $f^\star_{ld}$. Then we have
$
    f^\star_p \geq f^\star_l \geq f^\star_{ld}
$ by weak duality.
{Suppose that the running-averaged control policy in \eqref{eq:running average2} corresponds to an empirical primal objective value $\hat{f}_p(n)\geq f^\star_p$ for \eqref{eq:COHEMS} at iteration $n$,} and the averaged dual iterates
$\sum_{h=1}^H \lambda^{(h)}(t;n)/H$, $t=1,\ldots,T,$ correspond to an empirical dual value $\hat{f}_{ld}(n)\leq f^\star_{ld}$ for \eqref{eq:COHEMS stochastic dual}. {The normalized approximation gap between $\hat{f}_p(n)$ and $f^\star_p$
can be upper bounded as
\begin{align}\label{gap}
 \frac{\hat{f}_p(n)-f^\star_p}{f^\star_p}\leq \frac{\hat{f}_p(n)-\hat{f}_{ld}(n)}{\hat{f}_{ld}(n)}~\forall n,
\end{align}
where the right hand side term is the empirical duality gap which can be evaluated numerically and will be examined in Section \ref{subsec: convergence}.}
}
\end{Remark}

\vspace{-0.2cm}
\subsection{Extensions}\label{subsection extensions}
{In this subsection, we discuss two interesting extensions of the proposed CoHEM design.}

{\bf 1) Joint power procurement and CoHEM optimization}: As we mentioned in Remark 2, the proposed approach optimizes the CoHEM scheduling in an off-line manner. Therefore, it is possible to optimize the CoHEM scheduling for the next day and determine the power bid $\{P(t)\}_{t=1}^T$ jointly\footnote{If the hour-ahead market is available, then the retailer can also jointly determine the CoHEM scheduling and the power bid for the next hour. Here, we illustrate the joint design problem by assuming the day-ahead market only.}, provided that the aggregator can accurately estimate the locational marginal price (LMP) in the day-ahead wholesale market and the real-time prices $\{\pi_{\rm s}(t)\}_{t=1}^T$ and $\{\pi_{\rm p}(t)\}_{t=1}^T$ beforehand. Note that the two problems have different time scales -- the power procurement is in the day-ahead market where the retailer submits bids $B(1),\ldots,B(24),$ for 24 hours of the next day; while the CoHEM scheduling is for the real-time market where the unit of (discrete) time is usually in minutes. For ease of illustration, let us assume that sampling time interval in real time is 15 minutes. Then $P(t)=B(\lceil {t}/{4} \rceil)$ for $t=1,\ldots,96.$ Let
$C_{{\rm b},\ell}(B(\ell))$ denote the (convex) cost for the power bid at hour $\ell$, $\ell=1,\ldots,24$.
By adding $C_{{\rm b},\ell}(B(\ell))$ to the real-time cost in \eqref{eq:deviation cost}, the total cost of the retailer is given by

 {\small \begin{align}\label{eq:deviation cost + power bit}
 \!\!\!\!\text{Cost}=&\sum_{t=1}^{96} \left[ \pi_{\rm s}(t)\left(B(\lceil {t}/{4} \rceil)-\sum_{h=1}^H \E[L_{\rm total}^{(h)}(t)]\right)^+ \notag \right.\\
    &\left.+\pi_{\rm p}(t)\left(\sum_{h=1}^H E[L_{\rm total}^{(h)}(t)]-B(\lceil {t}/{4} \rceil)\right)^+ \right]\!\!+\!\sum_{\ell=1}^{24}C_{{\rm b},\ell}(B(\ell)),
\end{align}
 }\hspace{-0.19cm} and the associated joint power procurement and CoHEM scheduling problem can be formulated as
\begin{align}\label{joint design}
\!\!\!\!\!\!  \min_{\substack{B(\ell)\geq 0, \ell=1,\ldots,24,\\ s_{i,1}^{(h)},s_{i,2}^{(h)},\ldots \forall~i,h}}~&  \text{Cost~function~in~\eqref{eq:deviation cost + power bit}} \notag \\
  \text{s.t.}~&  \text{ \eqref{eq:COHEMS b} and constraints~in~\eqref{eq:COHEMS lowerbound c}~to~\eqref{eq:COHEMS lowerbound e}}.
  \end{align}
By following the same reformulation steps and dual decomposition technique in \eqref{eq:COHEMS lowerbound1} to \eqref{eq:COHEMS stochastic dual}, we can obtain the dual problem of \eqref{joint design} as
\begin{align}\label{joint design dual}
\max_{\substack{\lambda(t), \\t=1,\ldots,T}}~&\Bigg\{
\sum_{h=1}^H \Phi^{(h)}(\lambdab)  + \Psi(\lambdab) \Bigg\}  \notag \\
\text{s.t.}&~0\leq \lambda(t) \leq \pi_{\rm s}(t)+\pi_{\rm p}(t)~\forall t=1,\ldots,T,
\end{align}
where $\Phi^{(h)}(\lambdab)$, similar to (21), is given by

{ \begin{align} \label{inner minimization1}
  &\Phi^{(h)}(\lambdab)= \notag \\
  &\min_{s_{i,1}^{(h)},s_{i,2}^{(h)},\ldots}
  \sum_{t=1}^T(\lambda(t)-\pi_{\rm s}(t)) \left(\sum_{i=1}^N \E\!\left[D_i^{(h)}(t)\right]+U^{(h)}(t)\!\right) \notag \\
  &~~~~~~~~~~\text{s.t.}~\text{constraints~in~\eqref{eq:COHEMS lowerbound c}~to~\eqref{eq:COHEMS lowerbound e}},
\end{align}}\hspace{-0.19cm} and $\Psi(\lambdab)$ is given by
{ \begin{align}\label{power minimization}
  &\Psi(\lambdab)= \notag \\
  & \min_{\substack{B(\ell)\geq 0,\\ \ell=1,\ldots,24} } \sum_{\ell=1}^{24} C_{{\rm b},\ell}(B(\ell))-\sum_{t=1}^{96}(\lambda(t)-\pi_{\rm s}(t)) B(\lceil {t}/{4} \rceil).
\end{align}}\hspace{-0.19cm}
One can see from \eqref{joint design dual} that the optimization of $\Phi^{(h)}(\lambdab)$, $h=1,\ldots,H$, and $\Psi(\lambdab)$ are completely separable, and thus, the inner primal minimization of \eqref{joint design dual} can be carried out in a fully parallel manner. {In particular, \eqref{inner minimization1} can be solved by the MDP method in Section \ref{subsec: MDP for HEM}, while \eqref{power minimization} is a convex problem which can be solved by off-the-shelf convex solvers \cite{cvx}.} The only difference from \eqref{eq:COHEMS stochastic dual0} is that either an aggregator or one of the customers may need to be in charge of solving \eqref{power minimization}. Then the joint design problem \eqref{eq:deviation cost + power bit} can be handled in a similar decentralized fashion as Algorithm 1. {Simulation results to be presented in the next section will show that this joint power procurement and CoHEM scheduling can further reduce the overall cost of the retailer.}

{\bf 2) Extension to general convex cost functions:}
In some cases, the real-time cost function can be more complicated than that in \eqref{eq:deviation cost} \cite{LiLow2011PES} .
The proposed CoHEM design problem \eqref{eq:COHEMS} and Algorithm 1 can be extended to other general (convex, increasing) real-time cost functions. To illustrate this, let us rewrite \eqref{eq:COHEMS lowerbound} as follows
\begin{align}\label{COHEM with general cost}
\!\!\!\!\!\!  \min_{\substack{s_{i,1}^{(h)},s_{i,2}^{(h)},\ldots\\ \forall~i,h}}~&\sum_{t=1}^T \bigg\{C_{{\rm s},t}\bigg[\left(\hat{P}(t)-\sum_{h=1}^H \sum_{i=1}^N\E[D_i^{(h)}(t)]\right)^+\bigg] \notag \bigg. \\
    &~~~~~~\bigg.+C_{{\rm p},t}\bigg[\left(\sum_{h=1}^H \sum_{i=1}^N\E[D_i^{(h)}(t)]-\hat{P}(t)\right)^+\bigg] \bigg\}\notag \\
  \text{s.t.}~& \text{constraints~in~\eqref{eq:COHEMS lowerbound c}~to~\eqref{eq:COHEMS lowerbound e}},
\end{align}
where $C_{{\rm s},t}[\cdot]$ and $C_{{\rm p},t}[\cdot]$ denote the cost functions for buying additional power and absorbing extra power at time $t$, respectively; they are assumed to be convex and increasing. By introducing the two slack variables
\begin{align}
  z(t)&=\left(\sum_{h=1}^H \sum_{i=1}^N\E[D_i^{(h)}(t)]-\hat{P}(t)\right)^+~\forall t,\\
  y(t)&=\left(\hat{P}(t)-\sum_{h=1}^H \sum_{i=1}^N\E[D_i^{(h)}(t)]\right)^+~\forall t,
\end{align} one can write \eqref{COHEM with general cost} as
\begin{subequations}\label{COHEM general cost 2}
\begin{align}
\!\!\!\!\!\!  \min_{\substack{s_{i,1}^{(h)},s_{i,2}^{(h)},\ldots\\ \forall~i,h}}~&\sum_{t=1}^T \bigg\{C_{{\rm s},t}[z(t)] +C_{{\rm p},t}[y(t)] \bigg\} \\
  \text{s.t.}~&  \text{constraints~in~\eqref{eq:COHEMS lowerbound c}~to~\eqref{eq:COHEMS lowerbound e}}, \notag \\
  & z(t)\geq 0, ~y(t)\geq 0,~\forall t, \notag \\
  &z(t)\geq \sum_{h=1}^H \sum_{i=1}^N\E[D_i^{(h)}(t)]-\hat{P}(t)~\forall t, \label{COHEM general cost 2 C1}\\
  &y(t)\geq \hat{P}(t)-\sum_{h=1}^H \sum_{i=1}^N\E[D_i^{(h)}(t)]~\forall t. \label{COHEM general cost 2 C2}
 \end{align}\end{subequations}
{Similar to \eqref{eq:COHEMS stochastic dual0}, one can show that the dual problem of \eqref{COHEM general cost 2} has a separable inner minimization part, but has two additional minimization subproblems. These two subproblems can be handled by an aggregator or one of the voluntary residences. Moreover,
the dual variables associated with constraints \eqref{COHEM general cost 2 C1} and \eqref{COHEM general cost 2 C1} can be updated by the consensus projected subgradient method, analogous to Algorithm 1.} The detailed derivations are omitted here.

\vspace{-0.4cm}
\section{Simulation Results}\label{sec: simulations}
{Extensive simulation results are presented in this section to examine the performance of the proposed CoHEM architecture and Algorithm 1.}
\vspace{-0.2cm}
\subsection{Simulation Setting}\vspace{-0.0cm}
We consider a scenario where there are $H$ residential units, with 4 deferrable appliances in each residence. The optimization horizon is set to $96$ ($T=96$) which corresponds to a whole day with 24 hours and 4 quarters for each hour. The four controllable appliances considered are respectively washing machine, dish washer, tumble dryer and PHEV, all assumed to have a single operation mode and a single power profile for simplicity. The load profiles of the first three appliances are obtained according to the discrete load models in \cite[Table B1]{Paater2006}; the load profile of PHEV is set to be constant when on, with an instantaneous power consumption of 3 kW, and a working duration uniformly generated between 1 to 6 hours. We follow the synthetic method proposed in \cite{Paater2006} to generate the request probabilities $p_i^{(h)}(t)$ for the first three appliances in each house. For the PHEV, the request probability is set to 0.8 for three times that are uniformly distributed between 8 am and 12 pm, 5 pm and 0 am, and 0 am and 2 am, respectively.
The deadline constraints for the washing machine and dish washer are uniformly generated between 15 minutes to 2 hours, and for the tumble dryer and PHEV, they are generated uniformly between 15 minutes to 3 hours. The uncontrollable load $U^{(h)}(t)$ is contributed by the other 14 appliances listed in \cite[Table B1]{Paater2006} and is generated following the synthetic method in \cite{Paater2006}. We assumed that each residence $h$ can accurately estimate $U^{(h)}(t)$, for all $h$.

{If not mentioned specifically, the setting of Algorithm 1 is as follows. The initial value $\lambda^{(h)}(t;0)$ is set to $\frac{\pi_{\rm s}(t)+\pi_{\rm p}(t)}{2}$ for all $t=1,\ldots, 96,$ and $h=1,\ldots, H$; the step size $c_n$ is set to ${5}/(n+5)$; the number of averaging consensus step $\psi$ is set to $15$; 100 randomly generated realizations are used to estimate  $\sum_{i=1}^N\E\{D_i^{(h)}(t;n)\}$ in the Monte Carlo method.
Algorithm 1 is run for a predetermined number of iterations equal to 200.

The model predictive CEC method in \cite[Algorithm 1 \& Algorithm 2]{ChangPES12} is also simulated. The initial values, step size, and number of averaging consensus steps of \cite[Algorithm 2]{ChangPES12} are set to the same values as Algorithm 1. The maximum number of iterations of \cite[Algorithm 2]{ChangPES12} is set to 150. Note that, according to \cite[Algorithm 1]{ChangPES12}, \cite[Algorithm 2]{ChangPES12} has to be carried out 95 times, from time 1 to time 95.

For ease of elaboration, we set both prices $\pi_{\rm s}(t)$ and $\pi_{\rm p}(t)$ to one for all $t=1,\ldots,96$. In this case, the real-time cost in \eqref{eq:deviation cost} reduces to the total deviation between the aggregate load $\sum_{h=1}^H L_{\rm total}^{(h)}(t)$ and the day-ahead power purchase $\{P(t)\}_{t=1}^T$
\begin{align}\label{deviation cost}
	 \text{Deviation Cost}=\sum_{t=1}^T \bigg|P(t)-\sum_{h=1}^H L_{\rm total}^{(h)}(t) \bigg|.
\end{align}

The day-ahead bits $\{P(t)\}_{t=1}^T$ are generated as follows. Given the usage probabilities of customers, we use the Monte Carlo method to generate 50 realizations of aggregate unscheduled deferrable loads $\sum_{h=1}^H\sum_{i=1}^N{L}^{(h)}_i(t), t=1,\ldots,96$  (see \eqref{unscheduled load}), by which an estimate of $\sum_{h=1}^H\sum_{i=1}^N \E[{L}^{(h)}_i(t)]$, denoted by $\hat{L}(t)$, is obtained by taking the sample average. The day-ahead bid $\{P(t)\}_{t=1}^T$ used in the simulations is obtained by
\begin{align*}
   P(t)= \frac{1}{16}\sum_{\ell=1}^{16} \hat{L}( 16 (\lceil t/16 \rceil -1)+\ell)  + \sum_{h=1}^H U^{(h)}(t)
\end{align*}
for $t=1,\ldots, 96$. Note that in the first term of the above equation, we applied a peice-wise averaging for every 16-sample interval in order to emulate the effects of generation ramping constraints and that the retailer may have imperfect statistical information of customers and renewables\footnote{ The way we generate the day-ahead bid may not always be satisfactory from a practical point of view; however it is sufficient for one to assess and compare the scheduling capabilities of the developed DR algorithms. }.
}

\begin{figure}[!t] \centering
   \resizebox{0.45\textwidth}{!}{
     \psfrag{gamma}[Bl][Bl]{\huge $\gamma$ (dB)}
      \includegraphics{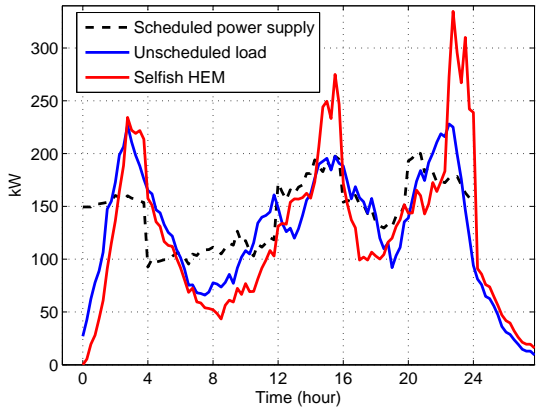}}
      \vspace{-0.2cm}
      \caption{Simulation results of the unscheduled power load and the power load scheduled by selfish HEM systems. {The number of residences $H$ is set to 100.}}
     \label{fig. HEM}\vspace{-0.3cm}
\end{figure}

\begin{figure}[!t]
\begin{center}
{\subfigure[][Model predictive CEC method in \cite{ChangPES12} ($H=100$) ]{\resizebox{.45\textwidth}{!}
{\includegraphics{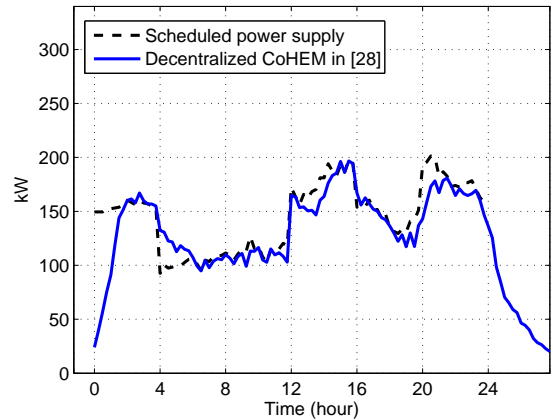}}}
}
\hspace{0.6pc}
{\subfigure[][{Algorithm 1 ($H=100$)}]{\resizebox{.45\textwidth}{!}{\includegraphics{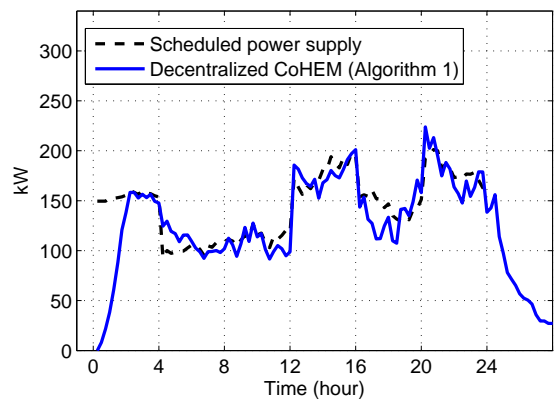}}}
 }
\hspace{0.6pc}
{\subfigure[][{Algorithm 1 ($H=400$)}]{\resizebox{.45\textwidth}{!}{\includegraphics{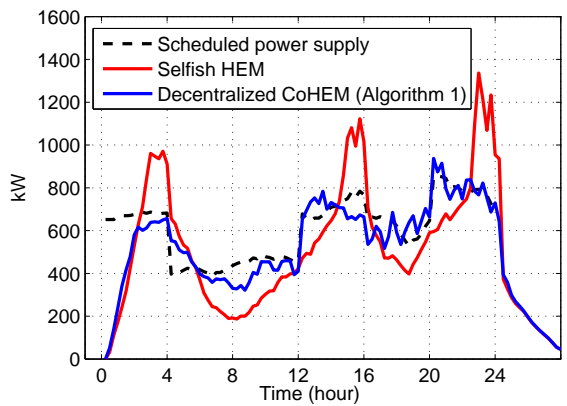}}}
 }

\end{center}\vspace{-0.3cm}
\caption{Simulation results of the loads scheduled by the model predictive CEC method in \cite{ChangPES12} and Algorithm 1.}
\vspace{-0.8cm}\label{fig. DCOHEM}
\end{figure}

\subsection{Selfish HEM v.s. CoHEM}
{Since there is no existing work that considers the same network utility and probabilistic models in this paper, we focus on comparing the proposed CoHEM architecture and Algorithm 1 with the selfish HEM system \eqref{eq:HEMS} and the model predictive CEC method in \cite{ChangPES12} .
}
Figure \ref{fig. HEM} shows the simulation results of the unscheduled load and the load scheduled by selfish HEM systems, {in the presence of 100 residential units ($H=100$). All the presented results are averaged over 50 randomly generated request realizations.}
In the simulation of selfish HEM, the residences are given a price signal that is inversely proportional to the power bid of the grid, i.e., $\pi(t)=1/P(t)$, in order to motivate the HEM systems in the residences to move their loads to the high-supply period.
One can see from Figure \ref{fig. HEM} that the selfish HEM design successfully moves the load to the high-supply region, but that also causes significant rebound peaks. {The aggregate deviation in \eqref{deviation cost} corresponding to the unscheduled load is $2823.3$ kW, but the deviation corresponding to selfish HEM increases to $4350.4$ kW. This result shows that the selfish HEM without coordination between the neighborhood would result in significant power imbalances and consequently considerable real-time cost to the retailer.
}

Figure \ref{fig. DCOHEM}(a) displays the power load scheduled by the model predictive CEC method in  \cite[Algorithm 1 \& Algorithm 2]{ChangPES12}.
By comparing Fig. \ref{fig. DCOHEM}(a) with Figure \ref{fig. HEM}, one can observe that the scheduled aggregate load by the model predictive CEC method \cite{ChangPES12} can follow the power supply, and the corresponding deviation cost dramatically decreases to $1549.8$ kW.
Figure \ref{fig. DCOHEM}(b) presents the power load scheduled by the proposed Algorithm 1.
{As one can see from this figure, Algorithm 1 performs comparably as the model predictive CEC method in \cite{ChangPES12}, but it has a slightly higher average deviation of {$2002.7$ kW}. Compared to the deviations of the unscheduled loads ($2823.3$ kW) and the selfish HEM system ($4350.4$ kW) shown in Fig. \ref{fig. HEM}, Algorithm 1 yields around {$29\%$ and $53\%$} reductions, respectively. We should emphasize again that Algorithm 1 has a much lower communication overhead than the model predictive CEC method in \cite{ChangPES12}. In particular, according to the setting described in Section IV-A, Algorithm 1 requires a total of $200 \times 15 =3000$ message exchanges; whereas the model predictive CEC method in \cite{ChangPES12} requires at most $150 \times 15 \times 95= 213750$ message exchanges. {
Figure \ref{fig. DCOHEM}(c) further shows the power loads scheduled by Algorithm 1 and the selfish HEM system in a neighborhood of 400 customers ($H=400$). Again, we see that the proposed CoHEM architecture and Algorithm 1 can significantly improve the power balancing and real-time cost of the aggregator.}

{To further look into how the number of residences affects the performance of the proposed CoHEM architecture, we list in Table I the normalized average deviation cost
for different numbers of residences. The results are obtained by testing the associated scheduling policy output by Algorithm 1 over 100 randomly generated request realizations. We see from Table I that there is a significant drop of the normalized deviation cost from $54.5$ kW to $22.4$ kW when the number of residences increase from 5 to 50; after $H\geq 50$, the normalized deviation costs remain relatively constant, showing that the performance of Algorithm 1 is quite robust against the size of the neighborhood.
}

\begin{table}[t]\caption{Normalized average deviation cost (in kW) versus number of residences.} \label{tab2} \centering
\begin{tabular}{|c|c|c|c|c|c|c|}
\hline $H$ & 5& 10 & 50 & 170 & 210 & 400 \\
\hline ${\rm Deviation~ cost} / H$  & {54.5}  & {46.4}  & {22.4}  & {21.4} &  {22.9} &  {20.6}
\\
\hline
\end{tabular}
\\
\vspace{0.2cm}
\hspace{-0.1cm}$^*$ The results are obtained by averaging over 100 simulation realizations. \end{table}

\begin{figure}[!t] \centering
   \resizebox{.45\textwidth}{!}{
      \includegraphics{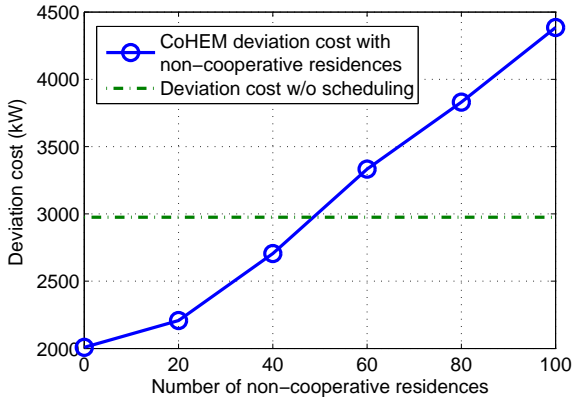}}
      \vspace{-0.0cm}
      \caption{Simulation results of deviation cost of Algorithm 1 in the presence of non-cooperative customers; {the simulation setting is the same as that for Fig. \ref{fig. DCOHEM}(b).}}
     \label{fig. noncooperative}\vspace{-0.5cm}
\end{figure}

\vspace{-0.3cm}
\subsection{Robustness of CoHEM} While all the residences in the neighborhood should cooperatively participate in the proposed CoHEM program, it is possible that there are some non-cooperative residences who selfishly keep using the selfish HEM policy. Here we examine how the number of such non-cooperative residences affect the performance of the proposed CoHEM architecture. Figure \ref{fig. noncooperative} presents the deviation cost in \eqref{deviation cost} for different number of non-cooperative residences.
The simulation setting is the same as Fig. \ref{fig. DCOHEM}(b). We can see from this figure that the deviation cost of the aggregator increases when there are more non-cooperative residences. However, compared to that without scheduling (the green dashed line), we observe that the aggregator can still make a profile out of the CoHEM program as long as there are more than {$50\%$} of residences in the neighborhood willing to follow the CoHEM scheduling policy.
This result demonstrates the robustness of the proposed CoHEM design against non-cooperative residences.

\vspace{-0.2cm}
\subsection{Convergence and Complexity}\label{subsec: convergence}\vspace{-0.0cm} In this subsection, we examine the convergence behavior of Algorithm 1 (see Remark 4) and its computation times. Figure \ref{fig. convergence}(a) displays the empirical objective value $\hat{f}_{ld}(n)$ of the dual problem \eqref{eq:COHEMS stochastic dual} versus the iteration number, for various numbers of residences in the neighborhood. Figure \ref{fig. convergence}(b) shows the corresponding normalized duality gap as discussed in \eqref{gap}.
{We can see from these figures that, within 150 iterations, the dual updates of Algorithm 1 as well as  the normalized duality gap converge asymptotically. The normalized duality gap shown in Figure \ref{fig. convergence}(b) shows that the gap between the empirical objective value $\hat{f}_{p}(n)$ of the original problem \eqref{eq:COHEMS} and $\hat{f}_{ld}(n)$ of the dual problem \eqref{eq:COHEMS stochastic dual} could be large; however, this does not necessarily imply that the CoHEM scheduling solution obtained from Algorithm 1 is far away from the optimal solution of problem \eqref{eq:COHEMS}, i.e., the gap between $\hat{f}_{p}(n)$ and ${f}_{p}^\star$ is not necessarily large. To further examine this aspect, we conduct a simulation where we set $H=2$ and $N=1$ (two residences and each of the residences has only one appliance). Under this setting, we are able to apply the MDP technique (as discussed in Section \ref{subsec: MDP for HEM}) to exhaustively find the optimal control policy for the CoHEM problem \eqref{eq:COHEMS} and the corresponding optimal objective value. In Fig. \ref{fig. convergence}(c), we plot $\hat{f}_{p}(n)$ and $\hat{f}_{ld}(n)$ of Algorithm 1 and also the optimal value of ${f}_{p}^\star$ obtained from the exhaustive MDP search.
Specifically, at iteration 200, we have $\hat{f}_{p}(200)=161.4$ and $\hat{f}_{ld}(200)=116.7$ while the optimal objective value is ${f}_{p}^\star=156.7$.
One can see from this figure that, although there is a large gap between $\hat{f}_{p}(n)$ and $\hat{f}_{ld}(n)$, $\hat{f}_{p}(n)$ is actually close to ${f}_{p}^\star$ (with a normalized accuracy $0.029$). 
While such inspiring result may not always hold true for large-scale problems (i.e., when $H$ and $N$ are large),
the evidenced results in Figures \ref{fig. HEM}, \ref{fig. DCOHEM} and \ref{fig. noncooperative} have demonstrated that Algorithm 1 is practically effective for large scale scenarios and can yield promising performance improvement for real-time power balancing.
}

Table II lists the computation times (in seconds) of Algorithm 1 for various numbers of residences in the neighborhood. The algorithm was run on the Matlab platform using a computer with a 4-core 2.6 GHz CPU and 12 GB RAM. Note that {while Algorithm 1 is a decentralized algorithm and the computations involved for solving \eqref{eq: stochastic subprob} and the Monte Carlo method can be parallelized, they can only be implemented sequentially in a computer.} The first row of Table II shows the average computation times per iteration (averaged over 400 iterations) ($T_{\rm c}/ {\rm ite}$) and the second row shows the computation times per iteration and per number of residences ($T_{\rm c}/ {\rm ite} / H$). It is interesting to see that $T_{\rm c}/ {\rm ite} / H$ remains relatively constant when $H$ increases, demonstrating that Algorithm 1 is truly scalable with the number of residences as long as a parallel computation can be implemented.

\begin{table}[t]\caption{Average computation time (in seconds) of Algorithm 1 versus number of residences.} \label{tab2} \centering
\begin{tabular}{|c|c|c|c|c|c|c|}
\hline $H$ & 50 & 90 &  170 & 210 & 400\\
\hline $T_{\rm c}/ {\rm ite}$ &  79.367 & 126.258 &  272.612 & 330.771 & 562.284 \\
\hline
\hline $T_{\rm c}/ {\rm ite} / H$  &   1.587  &  1.402  &  1.603 & 1.575 & 1.407\\
\hline
\end{tabular}
\\
\vspace{0.2cm}
\hspace{-0.1cm}$^*$$T_{\rm c}/ {\rm ite}$ stands for the average computation time per iteration and $T_{\rm c}/ {\rm ite} / H$ represents the computation time per iteration and per residential unit.
\end{table}

\begin{figure}[!t]
\begin{center}
{\subfigure[][Objective value $\hat{f}_{ld}(n)$ of \eqref{eq:COHEMS stochastic dual} ]{\resizebox{.45\textwidth}{!}{\includegraphics{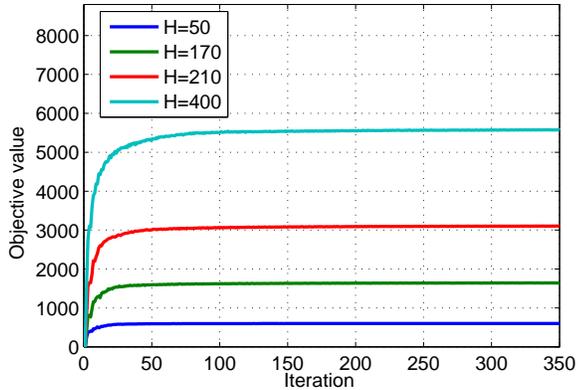}}}
}
\hspace{0.6pc}
{\subfigure[][Empirical normalized duality gap $(\hat{f}_p(n)-\hat{f}_{ld}(n))/\hat{f}_{ld}(n)$ ]{\resizebox{.45\textwidth}{!}{\includegraphics{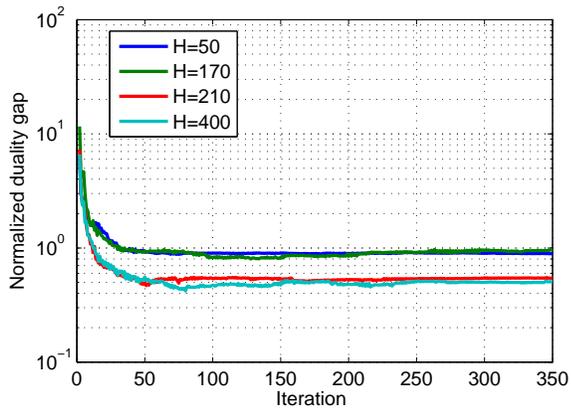}}}
 }
\hspace{0.6pc}
{\subfigure[][Empirical primal and dual objective values for $H=2$]{
\resizebox{.40\textwidth}{!}{\includegraphics{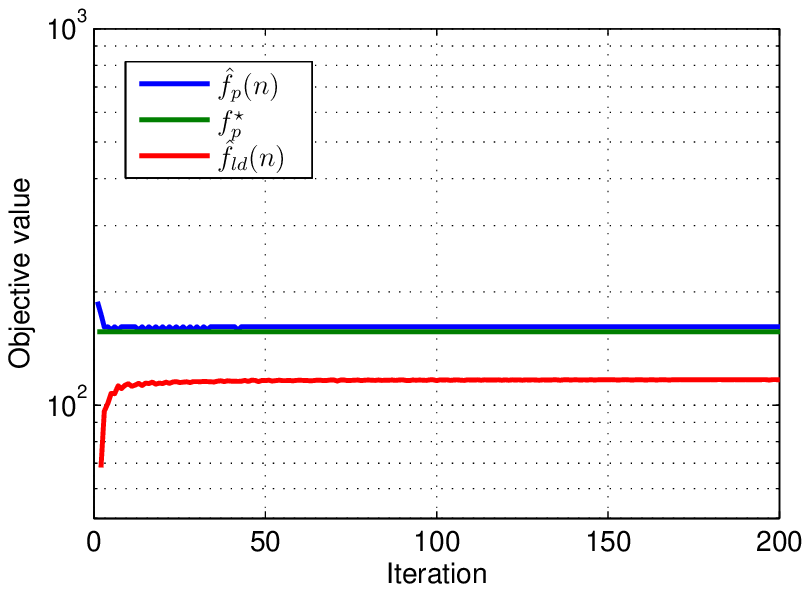}}}
}
\end{center}\vspace{-0.2cm}
\caption{Convergence curves of Algorithm 1.}
\vspace{-0.5cm}\label{fig. convergence}
\end{figure}

\vspace{-0.3cm}
\subsection{Joint power procurement and CoHEM scheduling}

In this subsection, we examine the performance of the joint power procurement and CoHEM scheduling formulation discussed in Section \ref{subsection extensions}. We consider the real-time deviation cost in \eqref{deviation cost} and the following quadratic cost function for power procurement
$$C_{{\rm b},\ell}(B(\ell))=B^2(\ell) \pi_{\rm LMP}(\ell),~\ell=1,\ldots,24,$$ where $\pi_{\rm LMP}(\ell)$ denotes the LMP for hour $\ell$. In the simulation, the LMP $\{\pi_{\rm LM}(\ell)\}_{\ell=1}^{24}$ are obtained from \url{https://www2.ameren.com/RetailEnergy/realtimeprices.aspx}, on day June 21, 2012. We put more weights on mitigating the power imbalance by considering the following weighed cost
\begin{align}\label{eq:deviation cost + power bit2}
 10\sum_{t=1}^{96} \bigg|B(\lceil {t}/{4} \rceil)&-\sum_{h=1}^H \E[ L_{\rm total}^{(h)}(t)] \bigg| +\sum_{\ell=1}^{24}B^2(\ell) \pi_{\rm LMP}(\ell).
\end{align}

\begin{figure}[!t]
\begin{center}
{\subfigure[][Algorithm 1 with a predetermined power bid $\{P(t)\}_{t=1}^T$. ]{\resizebox{.45\textwidth}{!}{\includegraphics{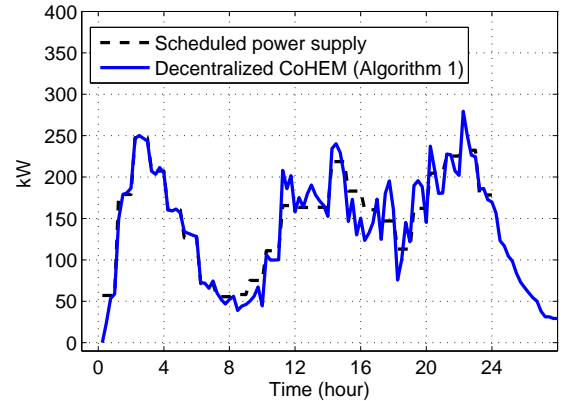}}}
}
\hspace{0.6pc}
{\subfigure[][Joint power procurement and CoHEM scheduling]{\resizebox{.45\textwidth}{!}{\includegraphics{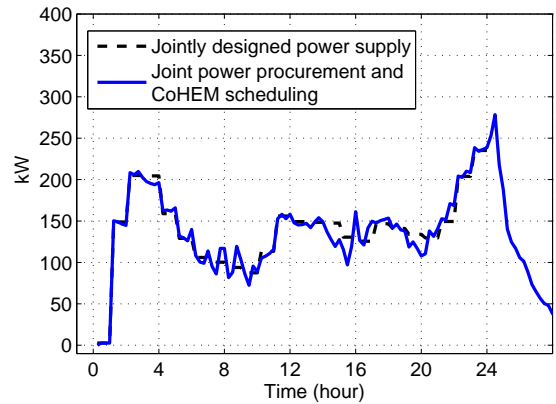}}}
 }
\end{center}\vspace{-0.1cm}
\caption{Performance of joint power procurement and CoHEM scheduling for a scenario with 130 residences ($H=130$).}
\vspace{-0.6cm}\label{fig. joint design}
\end{figure}

As a comparison with the joint power procurement and CoHEM scheduling design, we also simulate a scenario where the retailer first obtains a power bid $\{B(\ell)\}_{\ell=1}^{24}$ separately, followed by performing CoHEM scheduling (Algorithm 1) based on this power bid. In particular, the power bid is predetermined by minimizing the cost in \eqref{eq:deviation cost + power bit2} with $\sum_{h=1}^H \E[ L_{\rm total}^{(h)}(t)]$ replaced by the aggregate (unscheduled) power load $\sum_{h=1}^H \sum_{i=1}^N\E[ L_{i}^{(h)}(t)]$ (which can be estimated through the Monte Carlo method. The uncontrollable loads $U^{(h)}(t)$ are neglected here). The obtained power bid and the corresponding CoHEM scheduled load (averaged over 50 appliance request realizations) are presented in Fig. \ref{fig. joint design}(a), under the same simulation setting as that in Fig. \ref{fig. HEM}. The associated cost for power procurement $\sum_{\ell=1}^{24}B^2(\ell) \pi_{\rm LM}(\ell)$ and the average real-time deviation cost $\E[\sum_{t=1}^{96} |B(\lceil {t}/{4} \rceil)-\sum_{h=1}^H L_{\rm total}^{(h)}(t) |]$ are given by {$1685.0$ and $1874.4$}, respectively, leading to a total cost of {$3559.4$}. The jointly optimized power bid and the CoHEM scheduled load are shown in Fig. \ref{fig. joint design}(b).
The corresponding cost for power procurement $\sum_{\ell=1}^{24}B^2(\ell) \pi_{\rm LM}(\ell)$ and the average real-time deviation cost $\E[\sum_{t=1}^{96} |B(\lceil {t}/{4} \rceil)-\sum_{h=1}^H L_{\rm total}^{(h)}(t) |]$ are given by {$1111.0$ and $1427.4$}, respectively, which results in a lower total cost of {$2538.8$ (28.6\% reduction compared to $3559.4$ in Fig. \ref{fig. joint design}(a))}. We see that the joint design can yield lower costs for both real-time deviation and power procurement. We should emphasize here that the joint power procurement and CoHEM scheduling design requires accurate estimates for the LMP $\{\pi_{\rm LM}(\ell)\}_{\ell=1}^{24}$ in the day-ahead market as well as for the real-time prices $\{\pi_{\rm s}(t)\}_{t=1}^T$ and $\{\pi_{\rm p}(t)\}_{t=1}^T$ for the next whole day. Further investigations taking into account possible price estimation errors are needed in the future.

\vspace{-0.2cm}
\section{Conclusions and Future Directions}\label{sec: conclusions}
In the paper, we have presented a CoHEM architecture that coordinates the home energy scheduling of multiple residences in order to reduce the real-time power balancing cost. We first proposed a simple MDP approach for modeling the deferrable appliances and solving the individual HEM design problem. Then, we presented a decentralized algorithm (Algorithm 1) for handling the CoHEM design problem. The presented simulation results have demonstrated that the proposed CoHEM design as well as its decentralized algorithm can effectively decrease the real-time power balancing cost of the retailer.

{
In the future, we will extend the proposed load model and decentralized algorithms to thermostatically controlled appliances (e.g., heating, ventilating and air conditioning (HVAC) \cite{Du2011}). In particular, since HVAC has much shorter duty cycles compared to the non-interruptible loads, it can be scheduled myopically to reduce the uncertainty on $\hat P(t)$ due to imperfect information about the renewable energy sources and uncontrollable loads. In addition, it would be interesting to integrate the CoHEM with storage device control as well as distributed power generation control \cite{LavaeiLow2012}, and study this joint power flow control and CoHEM design problem form both economic and algorithmic aspects.
}

\vspace{-0.3cm}
\section{Acknowledgment}
\vspace{-0.1cm}
{The authors would like to sincerely thank the
anonymous reviewers whose comments have helped us improve
the manuscript significantly.}

\appendices

\vspace{-0.0cm}
\vspace{-0.0cm}
\vspace{-0.0cm} \footnotesize
\bibliography{smart_grid}

\end{document}